\documentclass[a4paper,11pt]{article}
\usepackage[DIV12]{typearea}
\usepackage{longtable}
\usepackage{booktabs,colortbl}
\usepackage{multirow}
\usepackage{amsmath}
\usepackage{amssymb}
\usepackage{bbm}
\usepackage[utf8]{inputenc}
\usepackage{pdflscape}
\usepackage{xspace}
\usepackage[caption=false]{subfig}
\usepackage{nicefrac}
\usepackage{graphicx}
\usepackage{cite}  
\usepackage[small]{caption2}
\usepackage{mathtools}
\usepackage{nccfoots}
\usepackage{tikz}
\usetikzlibrary{calc,positioning}

\usepackage[all,cmtip]{xy}
\newlength{\xywd}
\newcommand{\xyrightarrow}[2][]{%
  \sbox{0}{$\scriptstyle#1$}%
  \xywd=\wd0
  \sbox{0}{$\scriptstyle#2$}%
  \ifdim\wd0>\xywd \xywd=\wd0 \fi
  \xymatrix@C\dimexpr\xywd+1em\relax{{}\ar[r]^{#2}_{#1}&{}}%
}

\newcommand{\subind}[1]{{{\ensuremath\scriptscriptstyle{(\hspace{-0.7pt}#1\hspace{-0.7pt})}}}}
\newcommand{\rep}[1]{\ensuremath\boldsymbol{#1}}
\newcommand{\crep}[1]{\ensuremath\bar{\boldsymbol{#1}}}
\newcommand{\Z}[1]{\ensuremath{\mathbbm{Z}_{#1}}} 
\newcommand{\T}[1]{\ensuremath{\mathbbm{T}^{#1}}}

\newcommand{\SU}[1]{\ensuremath{\mathrm{SU}(#1)}}

\newcommand{\U}[1]{\ensuremath{\mathrm{U}(#1)}}
\newcommand{\E}[1]{\ensuremath{\mathrm{E}_{#1}}}
\newcommand{\e}{\mathrm{e}}
\newcommand{\I}{\mathrm{i}}
\newcommand{\Id}{\mathbbm{1}}

\newcommand{\x}{\ensuremath{\times}}
\newcommand{\vev}[1]{\ensuremath{\langle{#1}\rangle}}

\newcommand{\SL}[2]{\ensuremath{\mathrm{SL}(#1, #2)}}
\newcommand{\SG}{\ensuremath{\mathrm{S}}}
\newcommand{\TG}{\ensuremath{\mathrm{T}}}

\usepackage{xcolor}
\definecolor{darkgreen}{HTML}{109930}
\definecolor{pink}{rgb}{0.858, 0.188, 0.478}

\advance \headheight by 3.0truept       
\usepackage[pdftex,hidelinks,colorlinks=true,citecolor=darkgreen]{hyperref}
\hypersetup{
    pdftitle = {The eclectic flavor symmetries of T2/ZK orbifolds},
    pdfauthor = {Baur, Nilles, Ramos-Sanchez, Trautner, Vaudrevange}
}

\begin{document}

\begin{titlepage}

\begin{flushright}
\normalsize{TUM-HEP 1511/24}
\end{flushright}

\vspace*{1.0cm}

\begin{center}
{\Large\textbf{\boldmath The eclectic flavor symmetries of $\mathbbm{T}^2/\Z{K}$ orbifolds \unboldmath}}

\vspace{1cm}
\textbf{Alexander~Baur}$^{a,b}$,
\textbf{Hans~Peter~Nilles}$^{c}$,
\textbf{Sa\'ul Ramos--S\'anchez}$^{b}$,\\
\textbf{Andreas Trautner}$^{d}$ and 
\textbf{Patrick~K.S.~Vaudrevange}$^{a}$
\Footnote{*}{%
\href{mailto:alexander.baur@tum.de;nilles@th.physik.uni-bonn.de;ramos@fisica.unam.mx;trautner@mpi-hd.mpg.de;patrick.vaudrevange@tum.de}{\tt Electronic addresses} 
}
\\[5mm]
\textit{$^a$\small Physik Department, Technische Universit\"at M\"unchen,\\ James-Franck-Stra\ss e 1, 85748 Garching, Germany}
\\[2mm]
\textit{$^b$\small Instituto de F\'isica, Universidad Nacional Aut\'onoma de M\'exico,\\ POB 20-364, Cd.Mx. 01000, M\'exico}
\\[2mm]
\textit{$^c$\small Bethe Center for Theoretical Physics and Physikalisches Institut der Universit\"at Bonn,\\ Nussallee 12, 53115 Bonn, Germany}
\\[2mm]
\textit{$^d$\small Max-Planck-Institut f\"ur Kernphysik, \\ Saupfercheckweg 1, 69117 Heidelberg, Germany}
\end{center}

\vspace{1cm}

\vspace*{1.0cm}

\begin{abstract}
Only four $\mathbbm{T}^2/\Z{K}$ orbifold building blocks are admissible in heterotic string compactifications.
We investigate the flavor properties of all of these building blocks. In each case, we identify the traditional
and modular flavor symmetries, and determine the corresponding representations and (fractional) modular weights
of the available massless matter states. The resulting finite flavor symmetries include Abelian and non-Abelian
traditional symmetries, discrete $R$ symmetries, as well as the double-covered finite modular groups
$(S_3\x S_3)\rtimes\Z4$, $T'$, $2D_3$ and $S_3\x T'$. Our findings provide restrictions for bottom-up
model building with consistent ultraviolet embeddings.

\end{abstract}

\end{titlepage}

\newpage

\section{Introduction}

Modular symmetries open up new perspectives for flavor model building in particle 
physics. They can be derived from duality transformations in string theory (like exchanging 
winding and momentum string modes) and provide us with a consistent ultraviolet
(UV) theory as a well-defined starting point for top-down (TD) model building.

Up to now, however, most explicit model building has been performed within a 
bottom-up (BU) approach, see e.g.~\cite{Feruglio:2017spp,Kobayashi:2023zzc,Ding:2023htn}.
It requires various arbitrary choices, such as the selection
of the finite modular flavor group, the representations and the modular weights of the
matter fields. The few attempts within the top-down approach, on the other hand, 
show a very restrictive picture with specific choices of finite flavor groups,
representations and modular weights. We need a more general exploration of the
TD approach to identify the available possibilities and formulate rules that
should be relevant for further BU considerations. One of the goals of this paper
is to make contact between TD and BU constructions.

The modular symmetry group \SL{2}{\Z{}} underlying a finite modular flavor group
is connected to a two-torus $\mathbbm{T}^2$ 
suitably embedded in the six-dimensional compact space of string theory. The 
presence of chiral fermions in string compactifications can be achieved via an 
orbifold twist of $\mathbbm{T}^2$ and this leads us to four basic structures 
for finite modular flavor symmetries,
\begin{equation}
  \mathbbm T^2/\Z{K} \qquad\text{with}\quad K=2,3,4\ \text{and}\ 6\,,
\end{equation}
that could appear within the six extra compact dimensions of string theory. 
Such an embedding is easily achieved in the framework of orbifold 
compactifications of (heterotic) string theory that allows contact 
to realistic string constructions for the standard model of particle physics
with gauge group $\SU3\x\SU2\x\U1$ and three families of quarks and leptons. 
In a more general context, one could find modular symmetries in Calabi-Yau 
manifolds with elliptic fibrations. Twisted states and Yukawa-couplings
transform non-trivially under the modular group. An explicit example within 
the TD approach has been worked out for the $\mathbbm T^2/\Z3$ case~\cite{Baur:2021bly,Baur:2022hma}.

A full classification of the opportunities within the TD approach would
thus require as a next step a study of these four fundamental building blocks $\mathbbm T^2/\Z{K}$. 
With such an exploration we want to address the following questions:
\begin{itemize}
\item Which are the allowed finite modular groups?

\item Are they accompanied by other symmetries (known as
traditional flavor symmetries)?

\item What are the possible representations of the
matter fields under these groups?

\item  What are the allowed modular weights?

\item Is there a connection between these representations
and the modular weights?
\end{itemize}
Up to now, the known options from the TD approach are quite restrictive
and do not yet allow a contact to presently available BU constructions.

In the present paper we would like to analyze the four building blocks 
for modular symmetries in order to classify all the possibilities 
allowed in TD constructions. The \Z3 case is already well understood.
We shall review it in section~\ref{sec:T2overZ3} and provide the answers 
to the questions raised above.
Section~\ref{sec:T2overZ2} will be devoted to the \Z2 case. It is special
because it allows for two independent moduli of the torus, the K\"ahler 
modulus $T$ and the complex structure modulus\footnote{In the other three cases 
$K=3,4,6$ the complex structure modulus has to be fixed to special values to allow
for the \Z{K} twist.} $U$. Here we start with the modular group
$\SL{2}{\Z{}}_U\x\SL{2}{\Z{}}_T$, as will be analyzed in section~\ref{sec:T2overZ2}. 
This builds on earlier discussions~\cite{Baur:2020jwc,Baur:2021mtl}. There
has not yet been explicit TD model building in the \Z2 case.
Section~\ref{sec:T2overZ6} contains the first discussion of the \Z6 orbifold. 
The finite finite modular group is a direct product of the modular groups
of the \Z2 case ($S_3$) and the \Z3 case ($T'$) while the traditional flavor 
group is Abelian (\Z{36}). It remains to be seen what this might imply
for explicit model building. The same question arises in the \Z4 case,
which will be discussed in section~\ref{sec:T2overZ4}. Naively, we would
expect the modular group $S_4'$ (the double cover of $S_4$), but it 
turns out that the massless states of the string theory do not give 
faithful representations of this group but only of its subgroup 
$2D_3\cong\Z3\rtimes\Z4\cong[12,1]$, explored recently in BU models~\cite{Arriaga-Osante:2023wnu}.
Section~\ref{sec:lessons} will summarize our results and will formulate the lessons from TD model
building. We shall stress that modular flavor symmetry never comes alone, 
but will always be accompanied by a traditional flavor symmetry that is
important for explicit model building. We shall also emphasize the rule 
which states that there is a strict correspondence between a representation 
of the finite modular group and its allowed modular weight (a rule violated
in most of the BU constructions so far). In addition
we shall discuss the influence of these rules for model building. The \Z3 
case gives the most promising models from available TD considerations, 
both from the geometric point of view (three families from three twisted 
sectors) as well as the fact that here we have many promising realistic 
string constructions~\cite{Lebedev:2006kn,Nilles:2014owa,Carballo-Perez:2016ooy,Olguin-Trejo:2018wpw}.
In the other \Z{K} cases it would still be interesting to find out how 
three families of quarks and leptons fit into the representations of 
the flavor groups. Section~\ref{sec:conclusions} will discuss possible
generalizations of the present basic schemes to more moduli, such as the 
Siegel modular groups, see e.g.~\cite{Ding:2020zxw,Baur:2020yjl,Nilles:2021glx,Kikuchi:2023awe,Kikuchi:2023dow}.
This might also lead to situations that are beyond the orbifold compactification of string theory.

\section{\boldmath \texorpdfstring{$\mathbbm{T}^2/\Z{3}$}{T2/Z3} \unboldmath}
\label{sec:T2overZ3}

\begin{table}[t!]
\center
\begin{tabular}{|c|c||c|c|c|c|c|c|}
\hline
\multicolumn{2}{|c||}{nature}        & outer automorphism       & \multicolumn{5}{c|}{\multirow{2}{*}{flavor groups}} \\
\multicolumn{2}{|c||}{of symmetry}   & of Narain space group    & \multicolumn{5}{c|}{}\\
\hline
\hline
\parbox[t]{3mm}{\multirow{6}{*}{\rotatebox[origin=c]{90}{eclectic}}} &\multirow{2}{*}{modular}            & rotation $\mathrm{S}~\in~\SL{2}{\Z{}}_T$ & $\Z{4}$      & \multicolumn{3}{c|}{\multirow{2}{*}{$T'$}} &\multirow{6}{*}{$\Omega(2)$}\\
&                                    & rotation $\mathrm{T}~\in~\SL{2}{\Z{}}_T$ & $\Z{3}$      & \multicolumn{3}{c|}{}                      & \\
\cline{2-7}
&                                    & translation $\mathrm{A}$                & $\Z{3}$      & \multirow{2}{*}{$\Delta(27)$} & \multirow{3}{*}{$\Delta(54)$} & \multirow{4}{*}{$\Delta'(54,2,1)$} & \\
& traditional                        & translation $\mathrm{B}$                & $\Z{3}$      &                               & & & \\
\cline{3-5}
& flavor                             & rotation $\mathrm{C}=\mathrm{S}^2\in\SL{2}{\Z{}}_T$      & \multicolumn{2}{c|}{$\Z{2}^R$} & & & \\
\cline{3-6}
&                                    & rotation $\mathrm{R}=\gamma_\subind{3}\in\SL{2}{\Z{}}_U$ & \multicolumn{3}{c|}{$\Z{9}^R$}   & & \\
\hline
\end{tabular}
\caption{Eclectic flavor group $\Omega(2)$ for orbifolds $\mathbbm T^6/P$ that contain a 
$\mathbbm T^2/\Z{3}$ orbifold sector. In this case, $\SL{2}{\Z{}}_U$ of the complex
structure modulus is broken by $\langle U\rangle=\e^{\nicefrac{2\pi\I}{3}}$, resulting in a remnant
$\Z{9}^R$ $R$-symmetry. Including $\Z{9}^R$ enhances the traditional flavor group $\Delta(54)$ to 
$\Delta'(54,2,1)\cong [162,44]$ and, thereby, the eclectic group to 
$\Omega(2) \cong [1944, 3448]$. Note that $\Omega(1)\subset \Omega(2)$. Table from~\cite{Nilles:2020tdp}.}
\label{tab:Z3FlavorGroups}
\end{table}

Let us start with the $\mathbbm T^2/\Z3$ case as this has been studied previously
in detail. We summarize the results to set up the general
structure of our discussion. Geometry and string selection rules
lead to the traditional flavor symmetry $\Delta(54)$~\cite{Kobayashi:2006wq}. 
The finite modular symmetry is $\Gamma_3'\cong T'$, the
double cover of $\Gamma_3\cong A_4$~\cite{Baur:2019kwi,Baur:2019iai}. These results can be
directly determined from strings on the two-dimensional geometry of
$\mathbbm T^2/\Z3$. The embedding in the six-dimensional compact space
of string theory leads to an additional discrete $R$-symmetry $\Z9^R$~\cite{Nilles:2020tdp}.
The eclectic flavor group~\cite{Nilles:2020nnc} is the multiplicative closure
of these groups and turns out to be $\Omega(2)\cong[1944,3448]$
as shown in table~\ref{tab:Z3FlavorGroups}. Observe that this group does not have
$54\x24\x9 = 11,664$ elements, but only 1,944 due to two reasons: first, a
\Z3 non-$R$ subgroup of $\Z9^R$ coincides with the \Z3 point-group selection rule in $\Delta(54)$
and, second, a \Z2 $R$-symmetry within $\Delta(54)$ coincides with the \Z2 symmetry in $T'$ 
that characterizes the double cover related to the
\Z2 transformation $\mathrm S^2=-\Id_2$. This peculiar identification
originates from the properties of the outer automorphisms of
the Narain lattice as discussed\footnote{Traditional flavor groups $\Delta(54)$ and
$\Delta(27)$ lead to the same eclectic flavor group~\cite{Nilles:2020nnc}.} 
in ref.~\cite{Baur:2019iai,Nilles:2020nnc}.

Orbifold states come from untwisted and twisted sectors.
While the former typically correspond to (trivial or
nontrivial) singlet representations, twisted fields lead to
a more interesting situation. In the case under
consideration we have three fixed points and the twisted
sector fields transform as irreducible triplets of the
traditional flavor group $\Delta(54)$ as shown in
table~\ref{tab:Representations}. $\Delta(54)$ has two different 3-dimensional
representations, $\rep3_1$ and $\rep3_2$ (plus their complex conjugates),
and they correspond to massless string
states that do or do not contain twisted oscillators~\cite{Nilles:2020kgo}.
This three-dimensional structure gives an easy link to the
appearance of three families of quarks and leptons in
phenomenological applications. Under $T'$ the three twisted
fields do not transform as an irreducible triplet but as
a $\rep1\oplus\rep2'$ representation. The modular weights are
severely restricted and correlate with the representations
of $\Delta(54)$, $T'$ and $\Z9^R$ (with different
values for cases with and without twisted oscillators).
Each modular weight corresponds to a specific
representation of the eclectic flavor group. This leads
to severe restrictions which, in particular, forbid
so-called ``weighton'' fields (trivial singlets with various
non-trivial modular weights~\cite{King:2020qaj}).

Yukawa-couplings are modular forms of given weight $n$~\cite{Liu:2019khw}.
Since we are dealing with the double cover
$\Gamma_3' \cong T'$, $n$ is integer-valued
(in the case of $\Gamma_3 \cong A_4$, $n$ would be an even integer).
As we see from table~\ref{tab:Representations}, twisted fields have fractional modular
weights. This is allowed because of the presence of the
traditional flavor symmetry. The relevant co-cycle conditions
are fulfilled modulo a $\Delta(54)$ transformation.
Allowed terms in the superpotential automatically combine
the modular weights of the matter fields, such that they match
the integer modular weights of the Yukawa couplings.

The full scheme contains both the traditional as well as the
modular flavor symmetry. It is impossible to have one without the
other, and they both act on the same flavor space, not on direct product spaces.
In the present case this has important consequences
for the K\"ahler potential of the theory. It has previously
been pointed out~\cite{Chen:2019ewa} that the non-linearly realized nature
of the modular flavor group allows many terms in the
K\"ahler potential that could be problematic  for the
predictive power of the scheme. In the case under consideration,
however, the presence of the traditional flavor symmetry
$\Delta(54)$ forces the K\"ahler potential to its canonical
form~\cite{Nilles:2020kgo}, thus avoiding this potential problem.

For the \Z3 case, the connection to phenomenological
applications is straightforward. Three families of quarks
and leptons from the triplets of twisted fields seem to
be the best choice. There are many explicit string
constructions~\cite{Nilles:2014owa,Carballo-Perez:2016ooy,Olguin-Trejo:2018wpw}
that describe such a situation with
flavor groups $\Delta(54)$ and $T'$. Up to now, one
of these genuine top-down models has been worked out in
detail~\cite{Baur:2022hma}. More explicit constructions are needed to
fully explore the possibilities. This should include
constructions where matter fields correspond to states
with twisted oscillators. They here come in a different
triplet representation of the flavor group and different
modular weights (see table~\ref{tab:Representations}). This leads
to the peculiar situation that properties of the
low-energy effective action could distinguish between
matter fields that do or do not contain twisted
oscillators~\cite{Nilles:2020kgo}.

\begin{table}[t!]
\center
{
\begin{tabular}{cc|cc|cc}
\toprule
\multirow{2}{*}{sector} &string                & modular        & $T'\cong\Gamma'_3$ & $\Delta(54)$ & $\Z{9}^R$ \\
                        &state                 & weight         & irrep              & irrep        & charge    \\
\midrule
bulk                    &$\Phi_{\text{\tiny 0}}$    & $0$            & $\rep{1}$                & $\rep1$    & $0$ \\
                        &$\Phi_{\text{\tiny $-1$}}$ & $-1$           & $\rep{1}$                & $\rep1'$   & $3$ \\
first                   &$\Phi_{\nicefrac{-2}{3}}$  & $-\nicefrac23$ & $\rep{2}'\oplus\rep{1}$  & $\rep3_2$  & $1$ \\
                        &$\Phi_{\nicefrac{-5}{3}}$  & $-\nicefrac53$ & $\rep{2}'\oplus\rep{1}$  & $\rep3_1$  & $-2$ \\
second                  &$\Phi_{\nicefrac{-1}{3}}$  & $-\nicefrac13$ & $\rep{2}''\oplus\rep{1}$ & $\crep3_1$ & $2$ \\
                        &$\Phi_{\nicefrac{+2}{3}}$  & $+\nicefrac23$ & $\rep{2}''\oplus\rep{1}$ & $\crep3_2$ & $5$ \\
                        &$\Phi_{\nicefrac{-4}{3}}$  & $-\nicefrac43$ & $\rep{2}''\oplus\rep{1}$ & $\crep3_2$ & $-1$ \\
                        &$\Phi_{\nicefrac{+5}{3}}$  & $+\nicefrac53$ & $\rep{2}''\oplus\rep{1}$ & $\crep3_1$ & $-1$ \\
\midrule
                        &$Y^{(1)}_{\rep{2}''}(T)$   &  $1$           & $\rep{2}''$              & $\rep1$    & $0$ \\
\midrule
                        &$\mathcal W$               & $-1$           & $\rep{1}$                & $\rep1'$   & $3$ \\
\bottomrule
\end{tabular} 
}
\caption{Summary of flavor representations and charges for the bulk and two twisted sectors of the 
$\mathbbm{T}^2/\Z{3}$ orbifold. $T'$, $\Delta(54)$ and $\Z9^R$ combine nontrivially to the eclectic 
flavor group $\Omega(2) \cong [1944, 3448]$, as described in table~\ref{tab:Z3FlavorGroups}, 
cf.\ ref.~\cite{Baur:2021bly}.
\label{tab:Representations}}
\end{table}

\section{\boldmath \texorpdfstring{$\mathbbm{T}^2/\Z{2}$}{T2/Z2} \unboldmath}
\label{sec:T2overZ2}

\begin{table}[t!]
\center
{
\begin{tabular}{cc|cc|cc}
\toprule
\multirow{2}{*}{sector} &string                & modular weights & $[144,115]$ & $[32,49]$  & $\Z{4}^R$ \\
                        &state                 & $(n_T,n_U)$     & irrep       & irrep      & charge    \\
\midrule
bulk                    &$\Phi_{(0,0)}$        & $(0,0)$         & $\rep{1}_0$              & $\rep1_0$    & $0$ \\
                        &$\Phi_{(-1,-1)}$      & $(-1,-1)$       & $\rep{1}_0$              & $\rep1_0$    & $2$ \\
twisted                 &$\Phi_{(\nicefrac{-1}{2},\nicefrac{-1}{2})}$  & $(\nicefrac{-1}{2},\nicefrac{-1}{2})$ & $\rep4_1$  & $\rep4$  & $3$ \\
                        &$\Phi_{(\nicefrac{-3}{2},\nicefrac{1}{2})}$   & $(\nicefrac{-3}{2},\nicefrac{1}{2})$  & \multirow{2}{*}{$\rep4_1\oplus\rep4_1$}  
                                                                                                                   & $\rep4$  & $1$ \\
                        &$\Phi_{(\nicefrac{1}{2},\nicefrac{-3}{2})}$   & $(\nicefrac{1}{2},\nicefrac{-3}{2})$ &   & $\rep4$  & $1$ \\
\midrule
                        &$Y^{(2)}_{\rep{4}_3}(T,U)$ &  $(2,2)$       & $\rep{4}_3$              & $\rep1_0$  & $0$ \\
\midrule
                        &$\mathcal W$               & $(-1,-1)$      & $\rep{1}_0$              & $\rep1_0$  & $2$ \\
\bottomrule
\end{tabular} 
}
\caption{Summary of flavor representations and charges for the bulk and twisted sector of the 
$\mathbbm{T}^2/\Z{2}$ orbifold. The modular $(S_3^U\x S_3^T) \rtimes \Z4\cong[144,115]$ 
together with the traditional $(D_8\x D_8)/\Z2 \cong [32,49]$ and $\Z4^R$ flavor symmetries 
build an eclectic flavor group of order 4608, cf.\ ref.~\cite{Baur:2021mtl}.
\label{tab:Z2modularWeights}}
\end{table}

The \Z2 twist does not restrict the values of the moduli.
We thus have two unconstrained moduli, the $U$ (complex structure)
and $T$ (K\"ahler) moduli of the torus with modular group
$\SL{2}{\Z{}}_U\x\SL{2}{\Z{}}_T$. We obtain the traditional
flavor group $(D_8\x D_8)/\Z2 \cong [32,49]$ for generic
values of $T$ and $U$~\cite{Kobayashi:2006wq}. For specific values of
the moduli, this group is enhanced to [128,523] for the
``raviolo'' with $\vev{U}=\I$ and [96,204] for the tetrahedron with
$\vev{U}=\omega:=\e^{\nicefrac{2\pi\I}{3}}$~\cite{Baur:2020jwc,Baur:2021mtl}.
The finite modular flavor group is built on
$\Gamma_2'=\Gamma_2 \cong S_3$ for $U$ and $T$ separately and leads to
$(S_3^U\x S_3^T) \rtimes \Z4 \cong [144,115]$ where \Z4
contains the \Z2 mirror symmetry that maps $T$ to $U$ and
vice versa. The embedding in 10-dimensional string theory
leads to an additional $R$-symmetry $\Z4^R$ that combines
with the traditional flavor symmetry [32,49] to [64,266].
The eclectic flavor group has 4608 elements which leads
to a rich spectrum of (linearly realized) enhanced ``local
flavor unified groups'' (as displayed in figure 3 of~\cite{Baur:2021mtl}),
the largest of which is the union
of [1152,157463] with $\Z4^R$ that has 2,304 elements.

The representations of the matter fields are shown in
table~\ref{tab:Z2modularWeights}. The spectrum is severely restricted 
and displays a similar correlation between representations and
modular weights as found in the \Z3 case earlier.
Again we find fractional modular weights for the matter
fields, but this is unproblematic, because the modular
weights match with the integer values of the modular
forms $Y$ in allowed couplings.

The question concerning possible phenomenological
applications is still open. There are, however, no
irreducible 3-dimensional representations to describe
three families of quarks and leptons. Up to now,
in the \Z2 case, we do not have an explicit top-down
analysis of realistic string models with the spectrum
of the standard model of particle physics. The most
favorable scheme might be the consideration of models
with 4 families and one anti-family, where the former
could fit into the 4-dimensional representations of
[64,266] and [144,115]. More work is needed to clarify
this situation.

\section{\boldmath \texorpdfstring{$\mathbbm{T}^2/\Z{6}$}{T2/Z6} \unboldmath}
\label{sec:T2overZ6}

\begin{table}[h!]
\center
{
\begin{tabular}{cc|cc|cc}
\toprule
\multirow{2}{*}{sector} &string                & modular    & $S_3\x T'\cong\Gamma'_6$ & $\Z6^{\mathrm{(PG)}}$ & $\Z{36}^R$ \\
                        &state                 & weight     & irrep                    & PG selection rule     & charge    \\
\midrule
bulk          & $\Phi_0$                             & $0$            & $(\rep{1},\, \rep{1})$   & $0$ & $0$ \\
              & $\Phi_{-1}$                          & $-1$           & $(\rep{1},\, \rep{1})$   & $0$ & $6$ \\
$\vartheta$   & $\Phi_{1\mathrm{st}}$                & $-\nicefrac56$ & $(\rep{1}',\, \rep{1}')$ & $1$ & $1$ \\
$\vartheta^2$ & $\Phi_{2\mathrm{nd}}$                & $-\nicefrac23$ & $(\rep{1},\, \rep{2}')$  & $2$ & $2$ \\
$\vartheta^3$ & $\Phi_{3\mathrm{rd}}$                & $-\nicefrac12$ & $(\rep{2},\, \rep{1})$   & $3$ & $3$ \\
$\vartheta^4$ & $\Phi_{4\mathrm{th}}$                & $-\nicefrac13$ & $(\rep{1},\, \rep{2}'')$ & $4$ & $4$ \\
$\vartheta^5$ & $\Phi_{5\mathrm{th}}$                & $-\nicefrac16$ & $(\rep{1}',\,\rep{1}'')$ & $5$ & $5$ \\
\midrule
              & $Y^{(1)}_{(\rep{1},\,\rep{2}'')}(T)$ & $1$            & $(\rep{1},\, \rep{2}'')$ & $0$ & $0$ \\
              & $Y^{(1)}_{(\rep{2},\,\rep{2}')}(T)$  & $1$            & $(\rep{2},\, \rep{2}')$  & $0$ & $0$ \\
\midrule
              & $\mathcal W$                         & $-1$           & $(\rep{1},\, \rep{1})$   & $0$ & $6$ \\
\bottomrule
\end{tabular}
}
\caption{Summary of representations and charges of string states with no oscillator excitations in the $\mathbbm{T}^2/\Z{6}$ orbifold.
The only non-$R$ and non-modular discrete symmetry acting on matter states
corresponds to the standard $\Z6^{\mathrm{(PG)}}$ point-group selection rule.
In addition, $\Z{36}^R$ is the natural $R$-symmetry of the $\mathbbm{T}^2/\Z6$ orbifold, see e.g.~\cite{Nilles:2013lda,Schmitz:2014qoa}.
\label{tab:Z6summary}}
\end{table}

The \Z6 orbifold inherits properties both of the \Z2 and the
\Z3 case and we shall exploit this in our discussion. To allow
for the twist, we have to fix the complex structure modulus,
for example at $\vev{U}=\omega$, while the K\"ahler modulus $T$ remains
unconstrained. There is an untwisted and five twisted sectors
$\vartheta^k$ with $k=1,\ldots,5$. In the first twisted sector there
is only one fixed point and therefore the traditional flavor
symmetry must act Abelian here. In fact, as we will see in
more detail in the following, by computing the unbroken
remnants of the traditional flavor symmetries of the $\mathbbm{T}^2/\Z2$ 
and $\mathbbm{T}^2/\Z3$ sub-sectors of the $\mathbbm{T}^2/\Z6$ orbifold, 
one finds that the only unbroken non-$R$ symmetry is the $\Z6^{(\mathrm{PG})}$
symmetry from the point-group selection rule. In addition, there exists a
$\Z{36}^R$ $R$-symmetry of the \Z6-orbifold~\cite{Nilles:2013lda,Schmitz:2014qoa}, 
which contains $\Z6^{(\mathrm{PG})}$.
From the general rule given in appendix~\ref{sec:Gamma6'}, the finite modular
flavor group is a direct product of $\Gamma_2'\x\Gamma_3' \cong [144,128]$.
The eclectic flavor group is the union of $\Gamma_6'$ with $\Z{36}^R$
and has 2592 elements. This takes into account the fact that
$\Gamma_6'$ and $\Z{36}^R$ share a common $\Z2^R$ $R$-symmetry generated
by $\mathrm S^2$, leading to the eclectic flavor group\footnote{This relation
between $\Z{36}^R$ and $\Z{18}$ is similar
to the case of the $\mathbbm{T}^2/\Z3$ orbifold, where $\Delta(54)$ and
$\Delta(27)$ differ by a discrete $\Z2^R$ $R$-symmetry, see e.g.~\cite{Nilles:2024iqp}.}
$(\Gamma'_6\x\Z{36}^R)/\Z2^R\cong \Gamma'_6\x\Z{18}$.

Our strategy to determine the properties of the twisted fields
will be based on known results from the \Z2 case (for the $\vartheta^3$ sector)
and the \Z3 case (for the $\vartheta^2$ and $\vartheta^4$ sectors). Representations of the
first (and fifth) twisted sectors will then be determined by a careful
inspection of certain allowed Yukawa couplings. Our results
are summarized in table~\ref{tab:Z6summary}.

\subsection{\boldmath 3rd twisted sector \unboldmath}
\label{sec:Z2orbifoldstates}
The $\vartheta^3$-twisted sector of the $\mathbbm{T}^2/\Z6$ orbifold is
reminiscent of the $\vartheta$-sector of the
$\mathbbm{T}^2/\Z2$ orbifold with a fixed complex structure modulus
$\langle U \rangle=\omega$.
The modular weight of $n=-\nicefrac12$ of the twisted states
(in the absence of twisted oscillators) is inherited from the
\Z2 case and the relevant part of the finite modular flavor
group is $\Gamma_2'=\Gamma_2\cong S_3$.
The $\mathbbm{T}^2/\Z2$ orbifold has four fixed points
$\nicefrac{1}{2}(n_1e_1+n_2e_2)$ for
$n_1, n_2 \in \{0, 1\}$, where twisted string states are localized.
We denote them by
$\phi_{(n_1,n_2)}$ and consider them as a four-dimensional vector
$(\phi_{(0,0)},\phi_{(1,0)},\phi_{(0,1)},\phi_{(1,1)})^\mathrm{T}$ in flavor
space. The transformation properties under the modular
group have been given in ref.~\cite{Baur:2020jwc}
(see formulae (3.8a,b) and appendix C)
as well as ref.~\cite{Baur:2021mtl} (appendix D.3).

The $\SL{2}{\Z{}}_T$ transformations from ref.~\cite{Baur:2021mtl}
generate an $S_3$ finite modular group of the K\"ahler modulus.
To go from the $\mathbbm{T}^2/\Z2$ orbifold to the $\mathbbm{T}^2/\Z6$
orbifold, we have to consider $\Z6$-invariant linear combinations\footnote{%
A full discussion of this situation has to include the
so-called $\gamma$-phases, as explained in ref.~\cite{Nilles:2013lda}. 
For our discussion here, we concentrate on the states with $\gamma=0$.
\label{ref:footnote_gamma}}
of the $\mathbbm{T}^2/\Z2$ twisted matter states
$(\phi_{(0,0)},\phi_{(1,0)},\phi_{(0,1)},\phi_{(1,1)})^\mathrm{T}$.
We decompose the \Z6
orbifold action into $\Z3\x\Z2$. The \Z2-factor corresponds to
the \Z2 point group of $\mathbbm{T}^2/\Z2$ orbifold and, hence, acts trivially on
the $\T2/\Z2$ states.
On the other hand, the \Z3-factor interchanges the twisted matter states 
$\phi_{(1,0)}$, $\phi_{(0,1)}$ and $\phi_{(1,1)}$,
while it leaves $\phi_{(0,0)}$ (localized at the origin) invariant.
Hence, the \Z6 invariant states read
\begin{equation}
\Phi_{3\mathrm{rd}} ~=~ \begin{pmatrix}
\nicefrac{1}{\sqrt{3}}(\phi_{(1,0)}+\phi_{(0,1)}+\phi_{(1,1)})\\
\phi_{(0,0)}
\end{pmatrix}\;,
\end{equation}
while further orthogonal linear combinations of the $\T2/\Z2$ states are not 
invariant under the \Z6 orbifold action and are thus projected out.
From the behavior of $\phi_{(n_1,n_2)}$ under the modular $\mathrm{S}$ and 
$\mathrm{T}$ transformations associated with the K\"ahler modulus
$\hat{K}_{\mathrm{S}}$ and $\hat{K}_{\mathrm{T}}$ given in 
refs.~\cite{Baur:2020jwc,Baur:2021mtl}, we find that $\Phi_{3\mathrm{rd}}$ 
transforms as
\begin{equation}
\rho_{3\mathrm{rd}}(\hat{K}_{\mathrm{S}}) ~=~ \dfrac{1}{2} \begin{pmatrix}-1 & \sqrt{3} \\ \sqrt{3} & 1\end{pmatrix}
\qquad\mathrm{and}\qquad
\rho_{3\mathrm{rd}}(\hat{K}_{\mathrm{T}}) ~=~ \begin{pmatrix}1 & 0 \\ 0 & -1\end{pmatrix}\;.
\end{equation}
Taking into account the decomposition of $\Gamma'_6$ into its constituents $S_3\x T'$
(see appendix~\ref{sec:Gamma6'}), we find that the states from the third twisted sector
transform as a doublet $(\rep2,\rep1)$ of the finite modular
flavor group $\Gamma_6' \cong S_3\x T'$.

To obtain the traditional flavor symmetry in the $\vartheta^3$ sector of the $\mathbbm{T}^2/\Z6$ orbifold,
we start with the $\mathbbm{T}^2/\Z{2}$ orbifold with $(D_8 \x D_8)/\Z2$ as traditional flavor
symmetry at a generic point in moduli space. An explicit discussion of the generators $h_i$ of this group can be found
in appendix D.3.1 of ref.~\cite{Baur:2021mtl} and section 2 of ref.~\cite{Baur:2020jwc}.
Then, we analyze how the $\Z{6}$-invariant states $\Phi_{3\mathrm{rd}}$ transform under the 32
transformations of $(D_8 \x D_8)/\Z2$.
We find that the only unbroken traditional generator is the $\Z2$ transformation $h_5$, which
acts on $\Phi_{3\mathrm{rd}}$ with $\rho_{3\mathrm{rd}}(h_5)=-\Id_2$
and is associated with the point group selection rule.

In addition, there is a $\Z{36}^R$ $R$-symmetry associated with the $\mathbbm{T}^2/\Z6$ orbifold,
which in this sector can be understood as the residual geometric symmetry after the
complex structure modulus is fixed at $U=\omega$, see ref.~\cite[section 4.2]{Baur:2020jwc}.

\subsection{2nd twisted sector}
The 2nd (4th) twisted sector of the $\mathbbm{T}^2/\Z6$ orbifold is closely
related to the $\vartheta$ ($\vartheta^2$) sector of the $\mathbbm{T}^2/\Z3$ orbifold.
This implies that the modular weight of the twisted string
states (in the absence of twisted oscillator excitations) is
$-\nicefrac23$ for the second twisted sector and $-\nicefrac13$
for the fourth twisted sector. The modular transformations on the
three twisted string states $(X,Y,Z)^{\mathrm{T}}$
of the $\mathbbm{T}^2/\Z{3}$ orbifold have been given
explicitly in refs.~\cite{Baur:2019iai,Nilles:2020kgo}.
These transformations generate a $T'$ finite modular group of the K\"ahler
modulus. To go from the $\mathbbm{T}^2/\Z3$ orbifold to the
$\mathbbm{T}^2/\Z6$ orbifold, we have to build
$\Z6$-invariant linear combinations of the $\mathbbm{T}^2/\Z3$ twisted
matter states $(X,Y,Z)^{\mathrm{T}}$. Again, we decompose the $\mathbbm{T}^2/\Z6$ 
orbifold action into $\Z3\x\Z2$. 
The \Z3-factor corresponds to the \Z3 point group of $\mathbbm{T}^2/\Z3$
orbifold and, hence,
acts trivially on $(X,Y,Z)^\mathrm{T}$. On the other hand, the \Z2-factor
interchanges $Y$ and $Z$, while it leaves $X$, localized at the origin, 
invariant. Thus, in the simplest case of a trivial
$\gamma$-phase~\cite{Nilles:2013lda}, the \Z6-invariant states read
\begin{equation}
  \Phi_{2\mathrm{nd}} ~=~ \begin{pmatrix} \frac{1}{\sqrt{2}}(Y+Z) \\ -X \end{pmatrix}.
\end{equation}  
In contrast, the orthogonal state $\frac{1}{\sqrt{2}}(Y-Z)$
is non-invariant and, hence, not considered in the following.
Knowing from refs.~\cite{Baur:2019iai,Nilles:2020kgo} how $X$, $Y$, and $Z$ behave
under the $\mathrm{S}$ and $\mathrm{T}$ transformations of the K\"ahler modulus,
$\hat{K}_{\mathrm{S}}$ and $\hat{K}_{\mathrm{T}}$, we
find that $\Phi_{2\mathrm{nd}}$ transforms as
\begin{equation}
  \rho_{2\mathrm{nd}}(\hat{K}_\mathrm{S}) ~=~
    \dfrac{-\I}{\sqrt{3}}\begin{pmatrix}1&\sqrt{2}\\\sqrt{2}&-1\end{pmatrix}
  \qquad\text{and}\qquad
  \rho_{2\mathrm{nd}}(\hat{K}_\mathrm{T}) ~=~
    \begin{pmatrix}1&0\\0&\omega^2\end{pmatrix}.
\end{equation} 
Hence, using the properties of $T'$ within $\Gamma_6'$, as given in appendix~\ref{sec:Gamma6'},
we find that $\Phi_{2\mathrm{nd}}$ builds a $(\rep{1},\, \rep{2}')$
representation of the finite modular group $S_3\x T' \cong \Gamma'_6$ of
the $\mathbbm{T}^2/\Z6$ orbifold.

The $\mathbbm{T}^2/\Z{3}$ orbifold is endowed with a $\Delta(54)$ traditional flavor symmetry, 
as the completion of $\Delta(27)$ with a $\Z{2}^R$ $R$-symmetry. The action of $\Delta(54)$
on the $\mathbbm{T}^2/\Z{3}$ 
twisted matter states $\left(X,\, Y,\, Z \right)^\mathrm{T}$ is explicitly given in
section 2.2 of ref.~\cite{Nilles:2020kgo}.
In order to identify the unbroken traditional flavor symmetry in the $\vartheta^2$ ($\vartheta^4$) sector
of the $\mathbbm{T}^2/\Z6$ orbifold, we analyze how the \Z6-invariant states transform under
the 54 elements of $\Delta(54)$. We find that only the $\Z2^R$ $R$-symmetry
mentioned above and a $\Z3$ associated with the point-group selection rule are unbroken.
In addition, there exists a $\Z{9}^R$ $R$-symmetry in the $\mathbbm{T}^2/\Z{3}$ orbifold generated 
by a $\Z{3}$ sublattice rotation, see refs.~\cite{Nilles:2013lda,Nilles:2020gvu}.
This symmetry remains unbroken in the $\mathbbm{T}^2/\Z{6}$ orbifold
and contributes to the $\Z{36}^R$ $R$-symmetry.

\subsection{1st twisted sector}

In the first twisted sector of the $\mathbbm{T}^2/\Z{6}$ orbifold, there is a single localized 
string $\Phi_{1\mathrm{st}}$. In the absence of oscillators, its modular weight is $-\nicefrac56$. 
Being a single string in the first twisted sector, $\Phi_{1\mathrm{st}}$ has to form a 
one-dimensional representation of the finite modular group $S_3\x T'$.
While there exist in principle various one-dimensional representations of this group, only one of 
them leads to a non-vanishing superpotential that does not contain unphysical interactions for 
$T\rightarrow \I\infty$, as discussed in detail in appendix \ref{app:Z61st}.
This limit corresponds to decompactification and hence should not allow for interactions between states that 
are localized at different fixed points. For this to happen, the string state
$\Phi_{1\mathrm{st}}$ must transform as a $(\rep{1}',\, \rep{1}')$
representation of $S_3\x T' \cong \Gamma'_6$, such that
\begin{equation}
\rho_{1\mathrm{st}}(\hat{K}_\mathrm{S}) ~=~ -1
\qquad\text{and}\qquad
\rho_{1\mathrm{st}}(\hat{K}_\mathrm{T}) ~=~ \e^{2\pi\I\,5/6} \,.
\end{equation}
Besides the finite modular flavor symmetry, the string state of the first twisted sector is also
charged accordingly under the point group selection rule which completes the traditional flavor symmetry of the $\T2/\Z6$ orbifold.

\subsection{Superpotential}
Having determined the behavior of the string states under the flavor 
symmetries, let us briefly discuss possible interactions of the low 
energy effective field theory. To lowest (trilinear) order in twisted 
matter, the superpotential schematically reads
\begin{equation}\label{eq:Z6:Superpotential}
	\mathcal{W}  ~\supset~ 
	Y^{(1)}(T)\,\Phi_{1\mathrm{st}}\,\Phi_{2\mathrm{nd}}\,\Phi_{3\mathrm{rd}} ~+~ 
	Y^{(1)}(T)\,\Phi_{2\mathrm{nd}}^{(\mathrm{A})}\,\Phi_{2\mathrm{nd}}^{(\mathrm{B})}\,\Phi_{2\mathrm{nd}}^{(\mathrm{C})} ~+~
	Y^{(1)}(T)\,\Phi_{1\mathrm{st}}^{(\mathrm{A})}\,\Phi_{1\mathrm{st}}^{(\mathrm{B})}\,\Phi_{4\mathrm{th}}\,,
\end{equation}
where $Y^{(1)}(T)$ are weight $1$ modular forms of $\Gamma(6)$ given
in appendix \ref{sec:Gamma6Forms}. One can easily verify that these 
three terms are invariant under the point group selection rule, respect 
the $R$-symmetry, and yield an overall modular weight of $-1$ for 
$\mathcal{W}$. Moreover, all three terms contain couplings that are 
invariant under the non-Abelian finite modular flavor symmetry
$\Gamma_6'$. In detail, the symmetry invariant part of the first term 
of eq.~\eqref{eq:Z6:Superpotential} reads
\begin{align}\notag
	\mathcal{W} ~\supset&\,~ 
	~~Y^{(1)}_{(\rep{2},\,\rep{2}'),1}(T)~\Phi_{1\mathrm{st}}\,\Phi_{2\mathrm{nd},2}\,\Phi_{3\mathrm{rd},2}  
	~-~ Y^{(1)}_{(\rep{2},\,\rep{2}'),2}(T)~ \Phi_{1\mathrm{st}}\,\Phi_{2\mathrm{nd},1}\,\Phi_{3\mathrm{rd},2} \\
	& - Y^{(1)}_{(\rep{2},\,\rep{2}'),3}(T) ~\Phi_{1\mathrm{st}}\,\Phi_{2\mathrm{nd},2}\,\Phi_{3\mathrm{rd},1} 
	~+~ Y^{(1)}_{(\rep{2},\,\rep{2}'),4}(T) ~ \Phi_{1\mathrm{st}}\,\Phi_{2\mathrm{nd},1}\,\Phi_{3\mathrm{rd},1} \,,
\end{align}
where $\Phi_{j\mathrm{th},i}$ denotes the $i$-th component of a matter
field in the $j$-th twisted sector, while $Y^{(1)}_{(\rep{2},\,\rep{2}'),i}(T)$
refers to the $i$-th component of the $Y^{(1)}_{(\rep{2},\,\rep{2}')}(T)$ 
modular form.
For the second term of eq.~\eqref{eq:Z6:Superpotential} the symmetry
invariant couplings read in detail
\begin{align}\notag
	\mathcal{W} ~\supset&~~ Y^{(1)}_{(\rep{1},\,\rep{2}''),2}(T)\left(-\dfrac{1}{\sqrt{2}}\,\Phi_{2\mathrm{nd},1}^{(\mathrm{A})}\,\Phi_{2\mathrm{nd},1}^{(\mathrm{B})}\,\Phi_{2\mathrm{nd},1}^{(\mathrm{C})} 
	+ \Phi_{2\mathrm{nd},2}^{(\mathrm{A})}\,\Phi_{2\mathrm{nd},2}^{(\mathrm{B})}\,\Phi_{2\mathrm{nd},2}^{(\mathrm{C})}\right)
	\\\label{eq:SecondSuperpotTerm}
	& - \dfrac{Y^{(1)}_{(\rep{1},\,\rep{2}''),1}(T)}{\sqrt{2}}\left(\Phi_{2\mathrm{nd},1}^{(\mathrm{A})}\,\Phi_{2\mathrm{nd},1}^{(\mathrm{B})}\,\Phi_{2\mathrm{nd},2}^{(\mathrm{C})} 
	+\Phi_{2\mathrm{nd},1}^{(\mathrm{A})}\,\Phi_{2\mathrm{nd},2}^{(\mathrm{B})}\,\Phi_{2\mathrm{nd},1}^{(\mathrm{C})} 
	+\Phi_{2\mathrm{nd},2}^{(\mathrm{A})}\,\Phi_{2\mathrm{nd},1}^{(\mathrm{B})}\,\Phi_{2\mathrm{nd},1}^{(\mathrm{C})} \right)\;.
\end{align}
Note that this term is reminiscent of its related term in the superpotential 
of the $\T2/\Z3$ orbifold, in the sense that eq.~\eqref{eq:SecondSuperpotTerm} 
can be recovered from the $\T2/\Z3$ term given e.g.\ in \cite[eq.~(49)]{Nilles:2020kgo} 
when reduced to $\Z6$ invariant states. Recall also from appendix 
\ref{sec:Gamma6Forms} that the modular form $Y^{(1)}_{(\rep{1},\,\rep{2}'')}(T)$ 
of $\Gamma(6)$ is a modular form of $\Gamma(3)$ as well.
Finally, the third term of eq.~\eqref{eq:Z6:Superpotential} reads
\begin{equation}
	\mathcal{W} ~\supset~ 
	Y^{(1)}_{(\rep{1},\,\rep{2}''),1}(T) ~ \Phi_{1\mathrm{st}}^{(\mathrm{A})}\,\Phi_{1\mathrm{st}}^{(\mathrm{B})}\,\Phi_{4\mathrm{th},1}
	~+~
	Y^{(1)}_{(\rep{1},\,\rep{2}''),2}(T) ~ \Phi_{1\mathrm{st}}^{(\mathrm{A})}\,\Phi_{1\mathrm{st}}^{(\mathrm{B})}\,\Phi_{4\mathrm{th},2}\;.
\end{equation}

\subsection{\boldmath \texorpdfstring{$\Z6$}{Z6} Summary\unboldmath}
For the \Z6 case, TD phenomenological applications
have not yet been discussed. There seem to be many options:
we have various twisted sectors and a large finite modular
group $\Gamma_6'\cong S_3\x T'$. There are,
however, no irreducible triplet representations for 3 families
of quarks and leptons. The group $\Gamma_6'$ suggests
a $1 + 2$-pattern for families that could make the
connection to the \Z3 case with modular group $T'$.
There is, however, a significant difference between the
\Z3 and the \Z6 case when we compare the traditional
flavor groups. It is Abelian in the $\Z6$ case. In the
\Z3 case we have seen in ref.~\cite{Nilles:2020kgo} that the
non-Abelian traditional flavor group $\Delta(54)$ gave severe restrictions
on the K\"ahler potential that avoids the problems pointed
out in ref.~\cite{Chen:2019ewa}. It remains to be seen how this question will
influence explicit TD model building based on \Z6 orbifolds. At the moment
there is a large variety of choices including also the consideration
of states with twisted oscillators, which we have not yet
discussed in detail.

\section{\boldmath \texorpdfstring{$\mathbbm{T}^2/\Z4$}{T2/Z4} \unboldmath}
\label{sec:T2overZ4}

The traditional flavor symmetry is the semi-direct product of $\Z4^{(\text{PG})}\times\Z2$ resulting from the Abelianization
of the $\Z4$ space group~\cite[Table 1]{Ramos-Sanchez:2018edc} and a $\Z2$ stemming from a fractional lattice
translation. Explicitly, this yields the traditional flavor group~\cite{Kobayashi:2006wq}\footnote{%
The group $[16,13]$ is also known as the Pauli group and an alternative choice of the doublet generators are the three Pauli matrices.}
\begin{equation}
  (\Z4^{(\text{PG})}\x\Z2)\rtimes\Z2 ~\cong~ (D_8 \times \Z{4})/\Z{2} ~\cong~ [16, 13]\,.
\end{equation}
Further, embedding the $\mathbbm{T}^2/\Z4$ building block in a full six-dimensional string scheme
(such as $\mathbbm{T}^6/\Z{4}$, $\mathbbm{T}^6/\Z8\mathrm{-I}$ or $\mathbbm{T}^6/\Z{2}\times\Z{4}$,
see ref.~\cite{Fischer:2012qj}), leads to a
$\Z{16}^R$ discrete $R$-symmetry~\cite{Nilles:2020gvu}, which contains the non-$R$ subgroup $\Z4^{(\text{PG})}$
arising from the point group selection rule. Hence, the discrete $R$-symmetry enhances the traditional flavor symmetry
to $[16,13] \cup \Z{16}^R \cong [64,185]$.

The $\mathbbm{T}^2/\Z4$ orbifold exhibits an untwisted and three twisted sectors $\vartheta^k$, $k=1,2,3$.
There are three (two) inequivalent fixed points in the second (first and third) twisted sector, leading to
the same number of inequivalent string states, which build the flavor multiplets $\Phi_{2\mathrm{nd}}$
($\Phi_{1\mathrm{st}}$ and $\Phi_{3\mathrm{rd}}$). As in the previous case, we focus only on massless states with
trivial $\gamma$-phases and no twisted oscillator excitations, see also footnote~\ref{ref:footnote_gamma}.

On the other hand, to allow for the twist of the $\mathbbm{T}^2/\Z4$ orbifold, the complex structure
modulus is fixed at $\langle U \rangle=\I$ while the K\"ahler modulus $T$ is free, so that the modular
flavor symmetries emerge from the modular transformations on $T$.
In view of all of our previously discussed orbifolds, a naive extrapolation would suggest
the finite modular group of this setting to be $\Gamma_4'\cong S_4'\cong[48,30]$.\footnote{%
See~\cite{Novichkov:2020eep,Liu:2020akv,Abe:2023ilq} for recent discussions of this group in BU model building.}
However, we will see that the massless string states do not form faithful representations of the full possible finite modular group $S_4'$.
Instead, the group generated by the massless string spectrum and superpotential representations turns out to be
$[12,1]\cong 2D_3\cong S_4'/(\Z{2}\times\Z{2})$, which also happens to be a subgroup of $S_4'$.
We have checked that this statement holds true even allowing for twisted oscillators and non-trivial
$\gamma$-phases.\footnote{We conjecture that faithful representations of $S_4'$ can be found in the full massive string spectrum.}
In order to arrive at the flavor transformation properties of the twisted states, we will
again use the $\mathbbm{T}^2/\Z{2}$ case to deduce features of $\Phi_{2\mathrm{nd}}$ of
$\mathbbm{T}^2/\Z{4}$. In addition, the representations built by $\Phi_{1\mathrm{st}}$ (and $\Phi_{3\mathrm{rd}}$)
shall be determined by an explicit computation of the outer automorphisms of the Narain-lattice space group.
All modular and traditional charges are outlined in table~\ref{tab:Z4Charges}.

\begin{table}[t!]
\center
{
\begin{tabular}{cc|cc|ccc}
\toprule
\multirow{2}{*}{sector} & string                  &  modular     & $2D_3\cong[12,1]$               &  $[16,13]$  & $\Z{16}^R$ \\
                        & state(s)                &   weight     & irrep                           &  irrep      & charge \\
\midrule
bulk                    & $\Phi_0$                & $0$          & $\rep{1}$                       & $\rep{1}_0$ & $0$\\
                        & $\Phi_{-1}$             & $-1$         & $\rep{1}$                       & $\rep{1}_0$ & $4$\\

$\vartheta$             & $\Phi_{1\mathrm{st}}$   & $-\nicefrac{3}{4}$ & $\hat{\rep{2}}$           & $\rep{2}$   & $1$\\
$\vartheta^2$           & $\Phi_{2\mathrm{nd}}$   & $-\nicefrac{1}{2}$ & $\rep{2}\oplus\rep{1}'$   & $\rep{1}_6\oplus\rep{1}_2\oplus\rep{1}_5$ & $2$  \\
$\vartheta^3$           & $\Phi_{3\mathrm{rd}}$   & $-\nicefrac{1}{4}$ & $\hat{\rep{2}}$           & $\rep{\bar{2}}$ & $3$\\
\midrule
                        & $\mathcal{W}$           & $-1$         & $\rep{1}$                       & $\rep{1}_0$ & $4$\\
\bottomrule
\end{tabular}
}
\caption{Summary of flavor representations and charges in the $\mathbbm{T}^2/\Z{4}$ orbifold.
\label{tab:Z4Charges}}
\end{table}

The generators of the modular flavor symmetry $[12,1]\cong 2D_3$ are $\hat K_{\mathrm S}':= \mathrm{R}^4\hat K_{\mathrm S}$
and $\hat K_{\mathrm T}':= \mathrm{R}^{12}\hat K_{\mathrm T}$,
which satisfy the (double-cover) presentation (see eq.~\eqref{eq:Presentation2D3})
\begin{equation}
\label{eq:2D3inZ4}
2D_3 ~=~
\left\langle \hat K_{\mathrm S}',\,\hat K_{\mathrm T}' ~\Big|~
    \hat K_{\mathrm S}'^4 ~=~ \hat K_{\mathrm S}'^2\hat K_{\mathrm T}'^2 ~=~ (\hat K_{\mathrm S}'\hat K_{\mathrm T}')^3
                          ~=~ \mathrm{Id},~ \hat K_{\mathrm S}'^2\hat K_{\mathrm T}'~=~\hat K_{\mathrm T}'\hat K_{\mathrm S}'^2 \right\rangle\,.
\end{equation}
Combining the modular transformations $\hat K_{\mathrm S},\hat K_{\mathrm T}$ with $\mathrm{R}$
still yields a modular symmetry as the $\Z{16}^R$ generator $\mathrm{R}$ arises from the $\SL{2}{\Z{}}_U$
generator $\hat C_{\mathrm S}$ in the $\mathbbm T^2/\Z4$ orbifold~\cite{Nilles:2020gvu}. Interestingly,
$[12,1]$ is the only modular group that results from unfaithful representations of $S_4'$ and is
simultaneously a subgroup of the group generated by $\hat K_{\mathrm S},\hat K_{\mathrm T}$ and $\mathrm{R}$.

\subsection{2nd twisted sector}
The $\vartheta^2$-twisted sector of the $\mathbbm{T}^2/\Z4$ orbifold is
reminiscent of the $\vartheta$-sector of the
$\mathbbm{T}^2/\Z2$ orbifold with the complex structure modulus fixed at
$\langle U \rangle=\I$. The modular weights in the second twisted second sector 
are inherited from the \Z2 case and given by $n=-\nicefrac12$.
To obtain the second-twisted sector states of the $\mathbbm{T}^2/\Z4$ orbifold from the twisted matter states of the $\mathbbm{T}^2/\Z2$ orbifold we follow the same 
procedure as in the  $\mathbbm{T}^2/\Z6$ case described in section~\ref{sec:Z2orbifoldstates}.
In the present case, we need to form $\Z4$-invariant linear combinations of the $\mathbbm{T}^2/\Z2$
twisted matter states $(\phi_{(0,0)},\phi_{(1,0)},\phi_{(0,1)},\phi_{(1,1)})^\mathrm{T}$.
For trivial $\gamma$-phases, those are given by
\begin{equation}
\Phi_{2\mathrm{nd}} = \left(\phi_{(0,0)}, \frac{1}{\sqrt{2}}\left(\phi_{(1,0)} + \phi_{(0,1)}\right), \phi_{(1,1)}\right)^\mathrm{T}\;.
\end{equation}
The orthogonal state $\Phi_{2\mathrm{nd}}^-=\frac{1}{\sqrt{2}}\left(\phi_{(1,0)} - \phi_{(0,1)}\right)$ has $\gamma=-1$
and, hence, will be ignored here as we focus only on states with $\gamma=0$.

Performing the basis change
\begin{equation}
B_{2\mathrm{nd}}^{\text{trad}} ~=~
\begin{pmatrix}
\frac{1}{\sqrt2} & 0 & \frac{1}{\sqrt2} \\
0 & 1 & 0 \\
\frac{1}{\sqrt2} & 0 & -\frac{1}{\sqrt2}
\end{pmatrix}
\end{equation}
allows us to diagonalize the action on $\Phi_{2\mathrm{nd}}$ of the generators of the traditional
flavor symmetry $(\Z4^{(\text{PG})}\x\Z2)\rtimes\Z2\cong[16,13]$,
\begin{align}
\rho_{2\mathrm{nd}}(t_1)&= \begin{pmatrix}-1 & 0& 0 \\0 & -1 & 0\\0 & 0 & -1\end{pmatrix},&
\rho_{2\mathrm{nd}}(t_2)&= \begin{pmatrix}1 & 0& 0 \\0 & -1 & 0\\0 & 0 & 1\end{pmatrix},&
\rho_{2\mathrm{nd}}(t_3)&= \begin{pmatrix}1 & 0& 0 \\0 & 1 & 0\\0 & 0 & -1\end{pmatrix}.& \raisetag{30pt}
\end{align}
Hence, by comparing with the character table of the traditional flavor group
given in appendix~\ref{app:16-13}, we find that $\Phi_{2\mathrm{nd}}$ builds the representations
$\rep{1}_6\oplus\rep{1}_2\oplus\rep{1}_5$ of $[16,13]$.

On the other hand, to block-diagonalize the action of the finite modular symmetry,
we perform a basis change in flavor space using
\begin{equation}
B_{2\mathrm{nd}}^{\text{mod}} ~=~
\begin{pmatrix}
0 & \sqrt{\frac{2}{3}} & \frac{1}{\sqrt{3}} \\
1 & 0 & 0 \\
0 & \frac{1}{\sqrt{3}} & -\sqrt{\frac{2}{3}}
\end{pmatrix}\;,
\end{equation}
such that the action of the modular group on the basis-rotated $\Phi_{2\mathrm{nd}}$ is generated by 
\begin{align}
 &
 \rho_{2\mathrm{nd}}(\hat{K}_\mathrm{S}) = 
\begin{pmatrix}-\frac{1}{2} & \frac{\sqrt{3}}{2}&0\\\frac{\sqrt{3}}{2}&\frac{1}{2}&0\\0&0&1\\ \end{pmatrix}\;, &&\text{and}&&
\rho_{2\mathrm{nd}}(\hat{K}_\mathrm{T}) = 
\begin{pmatrix}1&0&0\\0&-1&0\\0&0&1\\ \end{pmatrix}\;.
& 
\end{align}
Notice that the representations of $\hat K_{\mathrm S}$ and $\hat K_{\mathrm T}$ satisfy
\begin{equation}
\rho_{2\mathrm{nd}}(\hat{K}_\mathrm{S})^2 ~=~ \rho_{2\mathrm{nd}}(\hat{K}_\mathrm{T})^2 ~=~ \left(\rho_{2\mathrm{nd}}(\hat{K}_\mathrm{S})\rho_{2\mathrm{nd}}(\hat{K}_\mathrm{T})\right)^3 ~=~ \Id_3\,,
\end{equation}
showing that $\Phi_{2\mathrm{nd}}$ does not transform faithfully under $\Gamma_4'$. Considering
$\rho_{2\text{nd}}(\mathrm R) = \e^{\nicefrac{4\pi\I}{16}}\Id_3$ and the definition of
$\hat K_{\mathrm S}'$ and $\hat K_{\mathrm T}'$ for eq.~\eqref{eq:2D3inZ4} and using the character
table given in section~\ref{sec:2D3}, we identify the irreducible representations of $\Phi_{2\mathrm{nd}}$
under $2D_3$ to be $\rep{2} \oplus \rep{1}'$, which are unfaithful and build only $S_3\cong 2D_3/\Z2$.

\subsection{1st twisted sector}
The representation of the generators of the traditional flavor symmetry
$(\Z4^{(\text{PG})}\x\Z2)\rtimes\Z2\cong[16,13]$ acting on the two states that build
the multiplet $\Phi_{1\mathrm{st}}$ is given by
\begin{align}
\rho_{1\mathrm{st}}(t_1)~&=~\begin{pmatrix}\I & 0\\0 & \I \end{pmatrix},&
\rho_{1\mathrm{st}}(t_2)~&=~\begin{pmatrix}1 & 0\\0 & -1\end{pmatrix},&
\rho_{1\mathrm{st}}(t_3)~&=~\begin{pmatrix}0 & 1\\1 & 0\end{pmatrix}.&
\end{align}
It then follows that matter states in the first twisted sector of the $\mathbbm T^2/\Z4$
orbifold build a $\rep2$ of $[16,13]$.

Further, the modular transformations of the first twisted sector can be derived using the geometrical
automorphisms of the Narain space group and the Operator Product Expansion (OPE) for twisted
strings~\cite{Lauer:1989ax,Lauer:1990tm}.
As discussed in detail in ref.~\cite{Baur:2019iai}, the modular flavor symmetries emerge from
the outer automorphisms of the Narain space group and we give a detailed derivation for the
present case in appendix~\ref{app:Z41stmodular}.
From a more BU perspective, one can also follow the discussion of~\cite{Nilles:2020nnc}
and identify the finite modular group based on the outer automorphisms of the full traditional
flavor symmetry $[64,185]$.
Both derivations agree to yield the generators\footnote{As usual the generators are unique
only up to multiplication with elements of the traditional flavor symmetry. Taking this into
account the results obtained here are fully compatible with ref.~\cite{Lauer:1990tm}, where
a different form of the $\mathrm{T}$ generator was stated.}
\begin{equation}
\label{eq:TrafoPhi1st_OPE}
	\rho_\mathrm{1st}(\hat K_\mathrm{S}) = \frac{1}{\sqrt2}
	\begin{pmatrix}
	1 & 1\\
	1 & -1
	\end{pmatrix}
	\qquad\text{and}\qquad
	\rho_\mathrm{1st}(\hat K_\mathrm{T}) =
	\begin{pmatrix}
	0 & \e^{\nicefrac{2\pi\I5}{8}}\\
	\e^{\nicefrac{2\pi\I3}{8}} &  0 
	\end{pmatrix}.
\end{equation}
Just as in the second twisted sector, taking $\rho_{1\text{st}}(\mathrm R) = \e^{\nicefrac{2\pi\I}{16}}\Id_2$,
we can define the generators $\hat K_{\mathrm S}'$ and  $\hat K_{\mathrm T}'$ as for eq.~\eqref{eq:2D3inZ4},
which allows us to identify that $\Phi_{1\text{st}}$ build the doublet $\hat{\rep2}$ representation of
$2D_3$.

The states of the third twisted sector of the  $\mathbbm T^2/\Z4$ orbifold must have the conjugate
quantum numbers of the states in the first twisted sector. Inspecting the character tables~\ref{tab:1613char}
and~\ref{tab:2D3char} of the traditional and modular flavor groups, we find that $\Phi_{3\text{rd}}$ builds
a $\bar{\rep2}$ representation of $[16,13]$ and a $\hat{\rep2}$ representation of $2D_3$.

\subsection{\boldmath \texorpdfstring{$\mathbbm{T}^2/\Z{4}$}{T2/Z4} summary\unboldmath}
Altogether, the finite modular group generated by the irreducible representations
of the fields and the superpotential is given by the binary dihedral group $2D_3\cong\Z3\rtimes\Z4\cong[12,1]$.
Our scheme thus provides a TD origin of a modular group that does not arise from a standard quotient of
\SL{2}{\Z{}} and a principal congruence group. In particular, the binary dihedral group was first identified
in ref.~\cite{Liu:2021gwa} as a finite modular group and was recently exploited in BU model building in ref.~\cite{Arriaga-Osante:2023wnu}.

Combining the modular flavor symmetry with the full traditional flavor symmetry $[64,185]$,
this furnishes the eclectic flavor symmetry of the $\mathbbm{T}^2/\Z{4}$ orbifold,
\begin{equation}
 [64,185]~\cup~2D_3~\cong~[384,5614]\,.
\end{equation}
Given the symmetries and representations listed in table~\ref{tab:Z4Charges}, this implies that
there are, to leading order, no allowed superpotential couplings. The existence of higher order
terms in the superpotential would have to be investigated taking into account also bulk and
winding modes, as well as massive string excitations.
This shows that the $\mathbbm{T}^2/\Z{4}$ compactification is doubly-special as compared to the
other cases discussed in this paper: First, it does not realize the full allowed finite modular
group $\Gamma_N'$. Second, the combined eclectic flavor symmetry prohibits 
lowest order superpotential couplings
altogether. Given the experience with the other orbifolds, this is rather unexpected and
investigating at what order superpotential terms are allowed is an interesting problem for future work.

\section{Top-down lessons for flavor model building}
\label{sec:lessons}

With our analysis, we have identified the possibilities for
flavor model constructions based on the four fundamental
building blocks from the twisted 2-torus. This allows some
general statements that might also be valid in more general
situations:

(i) The TD approach predicts the presence of a
traditional flavor symmetry. It is impossible to consider modular
flavor symmetry in isolation. This traditional flavor
symmetry can, in fact, be important for the control of
the K\"ahler potential of the theory. The relevant symmetry
to consider is the eclectic flavor symmetry, combining
the traditional and modular symmetries.

(ii) The choice of the finite modular flavor group is
very restricted to groups like $S_3$, $T'$, and
$2D_3$, but for example not $A_4$. To obtain larger
discrete groups, one might have to consider examples
beyond $\SL{2}{\Z{}}$ as a starting point. It is, however,
nice to see that promising groups like $T'$ and
$\Delta(54)$ appear within the eclectic picture as these
groups have been successfully used in many bottom-up (BU)
constructions.

(iii) The choice of the representations of the finite
modular flavor group is restricted as well. Only a few
of the representations are present in the TD constructions.
In fact, in all the cases we have not found a single
example of an irreducible 3-dimensional representation of
the modular flavor group (even if such a representation
was available). An illustrative example is $T'$
(which has an irreducible 3-dimensional representation),
where the twisted states transform as a $\rep1\oplus\rep2'$.
This example is particularly relevant for comparison with
BU constructions as there one often considers irreducible
3-dimensional representations to accommodate three families
of quarks and leptons as, for example, in  models based
on $A_4$ or $T'$. Here again, the traditional flavor
symmetry might come to rescue, for example with
irreducible 3-dimensional representations of $\Delta(54)$.
As a result of this comparison, the vast majority of
BU constructions are not compatible with the TD constructions
considered here.

(iv) There is no freedom to choose the modular weights of the
fields. They are ``locked'' with the representations of the
eclectic group. Once the representation is known, the modular
weight is essentially fixed. This is an important TD restriction, as in
many BU constructions the modular weights are freely chosen
to obtain successful fits. It also implies that there do not
exist so-called ``weighton'' fields with trivial transformation
properties under the eclectic group but different non-zero
modular weights. A careful choice of modular weights allows
for the appearance of so-called ``shaping'' symmetries within
BU constructions. Again this is not possible given the
TD restrictions.

(v) The traditional flavor symmetry leads to a control of the
K\"ahler potential. As the finite modular flavor symmetry is
mostly non-linearly realized, it allows for many unwanted
terms in the K\"ahler potential~\cite{Chen:2019ewa}. An 
illustrative example comes in the $\mathbbm{T}^2/\Z3$ case with modular
flavor group $T'$ and many additional parameters in the
K\"ahler potential. All of these are shown to be absent in the 
full theory because of the presence of the $\Delta(54)$ 
traditional flavor symmetry~\cite{Nilles:2020kgo}. It remains 
to be seen whether this control also exists in the $\mathbbm{T}^2/\Z4$ and
$\mathbbm{T}^2/\Z6$ cases, where the traditional flavor symmetries appear
to be less efficient (e.g.~Abelian in the case of $\mathbbm{T}^2/\Z6$).

(vi) Let us mention finally one aspect of string theory
models that has not yet been fully explored: the appearance
of states with twisted oscillators. Their modular weights
differ from those without twisted oscillators and they have
different $R$-charges. In the $\Z3$ case, such states transform
under a different irreducible 3-dimensional representation
($\rep3_1$ versus $\rep3_2$). They exhibit different allowed
couplings in the superpotential~\cite{Nilles:2020kgo}
that could be of
interest for explicit model building. So far all explicit
model constructions have concentrated on states without
twisted oscillators.

The statements made above came from the exploration of the
four simple 2-dimensional building blocks of the twisted
torus. Most of the rules survive once we go beyond these
simple cases, such as the strong restrictions on the allowed
representations and the ``locking'' of the modular weights
with these representations of the modular flavor group.
Bigger groups could possibly be obtained if one goes
beyond the modular group $\SL{2}{\Z{}}$. Some attempts have
been made with the Siegel modular group 
~\cite{Ding:2020zxw,Baur:2020yjl,Kikuchi:2023awe,Kikuchi:2023dow}
 or extensions to six dimensions and $\SL{2}{\Z{}}^3$ for
the product of three 2-tori~\cite{deMedeirosVarzielas:2019cyj,deMedeirosVarzielas:2021pug,Kikuchi:2023awm,King:2023snq}. 
Within the Siegel modular
group, there is one explicit example with ``modular group''
$S_4$~\cite{Ding:2020zxw}. 
It remains to be seen, how such examples can be
embedded in realistic string theory models~\cite{Nilles:2021glx}. 
Naively, there
seem to be many choices of symplectic symmetries with
many moduli, but one should keep in mind that we are
aiming at very specific string models with gauge group
$\SU3\x\SU2\x\U1$ and three families of
quarks and leptons and in most cases this is incompatible
with huge symplectic groups. In the case of heterotic
orbifolds with gauge groups like $\E8$ we need Wilson
lines to break the gauge group and they also lead to a
breakdown of the modular group. The models with $\SL{2}{\Z{}}$
and $\mathbbm T^2/\Z{K}$ building blocks, for example, are based on a
2-torus without a Wilson line in this particular torus. In any case, we need more explicit
string constructions to clarify these questions.

\section{Conclusions and Outlook}
\label{sec:conclusions}

The main challenge for flavor model building is the way
to incorporate three families of quarks and leptons. Given
the eclectic scheme, one has to select the representations
with respect to the traditional and modular flavor groups. In
general, one expects more predictive power with the use of
higher irreducible representations. Many of the
BU constructions assume quarks and, in particular leptons,
to be in 3-dimensional irreducible representations of the modular
group. In addition, one ignores the presence of a traditional
flavor symmetry, which is impossible in the UV complete constructions.
Given our results from the TD approach, we
see that we cannot make contact with essentially all of the
BU constructions on the market. This is already manifest
when we look at the available groups and the choices of
representations. One attractive class of BU models is based
on the $A_4$ modular group with matter representations 
in irreducible triplets of $A_4$. Unfortunately such a situation
cannot be found in our TD approach. Moreover, the discrepancy
between TD and BU constructions is most evident in the
choices of modular weights of the matter representations.
For BU models there appear many choices and this is an
important ingredient of model building, as it leads to
additional ``shaping'' symmetries. These choices of modular
weights are not available in the TD approach: modular
weights are fixed once we know the representations of the
modular group. In addition, the introduction of so-called
``weighton'' fields is not possible. We have shown, however,
that the TD approach predicts the presence of a traditional
flavor symmetry, mostly ignored in BU constructions. This
traditional flavor symmetry could be at the origin of ``shaping'' 
symmetries instead of the inappropriate use of modular weights.

On the positive side, we see the emergence of promising
flavor groups in the TD approach, most notably
$\Delta(54)$ for the traditional flavor group and
$T'$ for the modular one, groups that have been
used successfully in BU constructions. A strategy for
further BU model building could concentrate on these
groups $\Delta(54)$ and $T'$ including the
restrictions from TD considerations. In the TD approach,
up to now, only one example has been worked out in full
detail~\cite{Baur:2022hma}, but this is only the tip of the iceberg.
We would need to analyze more models in explicit string
constructions to exploit the full spectrum of
possibilities (including e.g.\ also the appearance of
fields with twisted oscillators).

\subsection*{Acknowledgments}
It is a pleasure to thank Xiang-Gan Liu and Michael Ratz for enlightening discussions.
A.B.\ was partially supported by the Deutsche Forschungsgemeinschaft (SFB1258).
A.B.\ and S.R.-S.\ are supported by UNAM-PAPIIT IN113223 and Marcos Moshinsky Foundation.
The authors are grateful to the Mainz Institute for Theoretical Physics (MITP) of the Cluster
of Excellence PRISMA$^+$ (project ID 390831469), for its hospitality and support during the completion of this work.

\newpage
\appendix

\section{\boldmath The modular group \texorpdfstring{$\Gamma'_6$}{Gamma6'} \unboldmath}
\label{sec:Gamma6'}

The group $\Gamma'_N := \Gamma/\Gamma(N)\cong\SL{2}{\mathbbm{Z}_N}$, can be decomposed into direct products~\cite{deAdelhartToorop:2011re}
\begin{equation}
\Gamma'_N ~=~ \prod_{p} \Gamma'_{p^{\lambda_p}}\;,
\end{equation}
given that $N$ is a product of primes, i.e.
\begin{equation}
N ~=~ \prod_{p} p^{\lambda_p}\;.
\end{equation}
Consequently, we get
\begin{equation}
\Gamma'_6 ~\cong~ \Gamma'_2 \x \Gamma'_3 ~\cong~ S_3 \x T' ~\cong~ [144, 128]\;.
\end{equation}
A presentation of this group is given by
\begin{equation}
\Gamma'_6 ~=~
\left\langle \SG,\,\TG ~\Big|~ \SG^4 ~=~ \TG^6 ~=~ (\SG\,\TG)^3 ~=~ \SG\,\TG^2\,\SG\,\TG^3\,\SG\,\TG^4\,\SG\,\TG^3 ~=~ \mathrm{Id} ~,~ \SG^2 \, \TG = \TG \, \SG^2 \right\rangle\;.
\end{equation}
It can be decomposed into its constituents as
\begin{equation}\label{eq:defS3}
S_3 ~=~ \left\langle 
\mathrm{S}_{S_3} := \SG\,\TG^3\,\SG^3, \, 
\mathrm{T}_{S_3} := \TG^3 ~\Big|~
\SG_{S_3}^2 ~=~ \TG_{S_3}^2 ~=~ (\SG_{S_3}\,\TG_{S_3})^3 ~=~ \mathrm{Id} \right\rangle \;,
\end{equation}
and
\begin{equation}
\label{eq:defTprime}
T' = \left\langle
\mathrm{S}_{T'} := \TG\,\SG\,\TG^4\,\SG^3\,\TG,\,\mathrm{T}_{T'} :=
\TG^4 ~\Big|~\SG_{T'}^4 = \TG_{T'}^3 = (\SG_{T'}\,\TG_{T'})^3 = \mathrm{Id},
~\SG_{T'}^2\,\TG_{T'}=\TG_{T'}\,\SG_{T'}^2
\right\rangle,
\end{equation}
where there is still the ambiguity that taking all elements of $S_3$ and multiplying them
with $\SG^2$ also yields a normal subgroup of $\Gamma'_6$ that commutes with the $T'$ normal subgroup.
The character tables of $S_3$ and $T'$ are given in tables \ref{tab:S3char} and \ref{tab:Tpchar}, respectively.

\begin{table}[h!]
	\centering
	\begin{tabular}{l|rrr}
		\toprule
		$S_3$ &  $[\mathrm{Id}]$ & $[\SG_{S_3}]$ & $[\SG_{S_3}\,\TG_{S_3}]$ \\
		\midrule
		$\rep{1}$     & $1$ & $1$  & $1$ \\
		$\rep{1}'$     & $1$ & $-1$  & $1$\\
		$\rep{2}$       & $2$ & $0$  & $-1$  \\
		\bottomrule
	\end{tabular}
	\caption{Character table of $S_3$.}
	\label{tab:S3char}
\end{table}

\begin{table}[!t]
	\begin{center}
		\begin{tabular}{l|rrrrrrr}
			\toprule
			$\mathrm{T}'$ & $\big[\mathrm{Id}\big]$ & $\big[\SG_{T'}^2\big]$ & $\big[\TG_{T'}\big]$ & $\big[\TG_{T'}^2\big]$ & $\big[\SG_{T'}\big]$ & $\big[\SG_{T'}^2\,\TG_{T'}\big]$ & $\big[\SG_{T'}^2\,\TG_{T'}^2\big]$ \\\midrule
			$\rep{1}$     & $1$ & $1$  & $1\phantom{{}^2}$                         	 & $1\phantom{{}^2}$      & $1$ & $1\phantom{{}^2}$ & $1\phantom{{}^2}$ \\
			$\rep{1}'$    & $1$ & $1$  & $\omega\phantom{{}^2}$                      & $\omega^2$ & $1$ & $\omega\phantom{{}^2}$ & $\omega^2$ \\
			$\rep{1}''$   & $1$ & $1$  & $\omega^{2}$ 	 							 & $\omega\phantom{{}^2}$ & $1$ & $\omega^2$ & $\omega\phantom{{}^2}$ \\
			$\rep{2}$     & $2$ & $-2$ & $-1\phantom{{}^2}$                          & $-1\phantom{{}^2}$ & $0$ & $1\phantom{{}^2}$ & $1\phantom{{}^2}$ \\
			$\rep{2}'$    & $2$ & $-2$ & $-\omega\phantom{{}^2}$                     & $-\omega^2$ & $0$ & $\omega\phantom{{}^2}$ & $\omega^2$ \\
			$\rep{2}''$   & $2$ & $-2$ & $-\omega^{2}$  							 & $-\omega\phantom{{}^2}$ & $0$ & $\omega^2$ & $\omega\phantom{{}^2}$ \\
			$\rep{3}$     & $3$ & $3$  & $0\phantom{{}^2}$                           & $0\phantom{{}^2}$ & $-1$ & $0\phantom{{}^2}$ & $0\phantom{{}^2}$ \\
			\bottomrule		
		\end{tabular}
		\caption{Character table of the finite modular group $\mathrm{T}' \cong \mathrm{SL}(2,3)$. Here, 
			$\omega:=\mathrm{e}^{\nicefrac{2\pi\I}{3}}$.}
		\label{tab:Tpchar}
	\end{center}
\end{table}

\subsection{\boldmath Relevant tensor products of \texorpdfstring{$S_3$}{S3} and \texorpdfstring{$T'$}{T'} \unboldmath}
\label{app:TensorProducts}
In order to discuss the invariants of a $\mathbbm{T}^2/\Z6$ orbifold yielding a $\Gamma_6'\cong S_3\x T'$
modular flavor symmetry, we provide first all nontrivial tensor products and Clebsch-Gordan coefficients of $S_3$.
In the basis of ref.~\cite{Ishimori:2010au}, we use
\begin{subequations}
\label{eq:S3ClebschGordans}
	\begin{align}
	\begin{pmatrix}x_1\\x_2 \end{pmatrix}_{\rep{2}} \otimes \begin{pmatrix}y_1\\y_2 \end{pmatrix}_{\rep{2}}
	     ~&=~ (x_1\,y_1 + x_2\,y_2)_{\rep{1}} \oplus (x_1\,y_2 - x_2\,y_1)_{\rep{1}'} \oplus \begin{pmatrix}x_1\,y_2+x_2\,y_1\\x_1\,y_1 - x_2\,y_2\end{pmatrix}_{\rep{2}}\,, \\
	(x)_{\rep{1}'} ~\otimes \begin{pmatrix} y_1 \\ y_2 \end{pmatrix}_{\rep{2}}
	     ~&=~ \begin{pmatrix} -y_2\,x \\ y_1\,x\end{pmatrix}_{\rep{2}}\,.
	\end{align}
\end{subequations}

Furthermore, we list relevant tensor products and Clebsch-Gordan coefficients of $T'$ in
the basis of ref.~\cite[with $p=\I,p_1=-p_2=1$]{Ishimori:2010au}. In appendix~\ref{app:Z61st} we
shall only need
\begin{equation}
\label{eq:TpClebschGordans}
\begin{pmatrix}x_1\\x_2 \end{pmatrix}_{\rep{2}'} \otimes \begin{pmatrix}y_1\\y_2 \end{pmatrix}_{\rep{2}'}
      ~=~ (x_1\,y_2 - x_2\,y_1)_{\rep{1}''} ~\oplus~ \begin{pmatrix}-x_1y_1\\x_2y_2\\ \frac{-1}{\sqrt2}(x_1y_2+x_2y_1) \end{pmatrix}_{\rep{3}}.
\end{equation}
and the fact that $\rep{1}''\otimes\rep{1}'=\rep{1}$, which can already be read off from table~\ref{tab:Tpchar}.

\subsection{\boldmath Modular forms of \texorpdfstring{$\Gamma(6)$}{Gamma6} \unboldmath}
\label{sec:Gamma6Forms}
A $q$-expansion for the modular forms of $\Gamma(6)$ with weight $1$ can be conveniently obtained from the computer algebra program
SageMath \cite{sagemath} and reads
\begin{subequations}\label{eq:SAGEmodularforms}
	\begin{align}
	\tilde y_1(T) ~&=~ 1 + 6\,q^{12} + 6\,q^{36} + \mathcal{O}(q^{37})\;,\\
	\tilde y_2(T) ~&=~ q + 2\,q^7 + 2\,q^{13} + 2\,q^{19} + q^{25} + 2\,q^{31}  + \mathcal{O}(q^{37})\;, \\
	\tilde y_3(T) ~&=~ q^2 + q^8 + 2\,q^{14} + 2\,q^{26} + q^{32}  + \mathcal{O}(q^{37})\;, \\
	\tilde y_4(T) ~&=~ q^3 + q^9 + 2\,q^{21} + q^{27}  + \mathcal{O}(q^{37})\;, \\
	\tilde y_5(T) ~&=~ q^4 + q^{16} + 2\,q^{28}  + \mathcal{O}(q^{37})\;, \\
	\tilde y_6(T) ~&=~q^6 - q^{12} + q^{18} + q^{24} - q^{36}  + \mathcal{O}(q^{37})\;,
	\end{align}
\end{subequations}
with $q:=\e^{\nicefrac{2\pi\I T}{6}}$. 
Alternatively, the following analytical expressions for the modular forms 
have been given in ref.\ \cite{Li:2021buv}:
\begin{subequations}\label{eq:analyticalmodularforms}
	\begin{align}
	\hat y_1(T) ~&=~ \dfrac{\eta^3(3\,T)}{\eta(T)}\;, &
	\hat y_2(T) ~&=~ \dfrac{\eta^3(T/3)}{\eta(T)}\;, &
	\hat y_3(T) ~&=~ \dfrac{\eta^3(6\,T)}{\eta(2\,T)}\;, \\
	\hat y_4(T) ~&=~ \dfrac{\eta^3(T/\;\!6)}{\eta(T/\;\!2)}\;, &
	\hat y_5(T) ~&=~ \dfrac{\eta^3(2\,T/\;\!3)}{\eta(2\,T)}\; &
	\hat y_6(T) ~&=~ \dfrac{\eta^3(3\,T/\;\!2)}{\eta(T/\;\!2)}\;,
	\end{align}
\end{subequations}
where $\eta(T)$ refers to the Dedekind eta function. 
Let us now decompose the modular forms into irreducible representations 
of $\Gamma'_6$ which reads
\begin{subequations}\label{eq:modularformsZ6}
	\begin{align}
	Y^{(1)}_{(\rep{1},\,\rep{2}'')}(T) ~&=~ 
	\begin{pmatrix}-3\sqrt{2}~\tilde y_3(T)\\ \tilde y_1(T) + 6\,\tilde y_6(T)\end{pmatrix}
	~=~ 
	\begin{pmatrix}-3\sqrt{2}~\hat y_1(T)\\3~\hat y_1(T) + \hat y_2(T)\end{pmatrix}
	\;, \\
	Y^{(1)}_{(\rep{2},\,\rep{2}')}(T) ~&=~ 
	-\sqrt{37}\begin{pmatrix}\tilde y_1(T)\\3\sqrt{2}~\tilde y_5(T)\\ -2\sqrt{3}~\tilde y_4(T) \\ -\sqrt{6}~\tilde y_2(T)\end{pmatrix}
	~=~  
	-\sqrt{37}\begin{pmatrix}3\,\hat y_3(T) + \hat y_5(T)\\ 3\sqrt{2}~\hat y_3(T)\\ \frac{1}{\sqrt{3}}(3\hat y_3(T) - \hat y_4(T) + \hat y_5(T) -3\hat y_6(T))\\ \sqrt{6}\,\hat y_3(T) - \sqrt{6}\,\hat y_6(T)\end{pmatrix}
	\;.
	\end{align}
\end{subequations}
Under a modular transformation, the first form 
$Y^{(1)}_{(\rep{1},\,\rep{2}'')}(T)$ transforms as
\begin{equation}
Y^{(1)}_{(\rep{1},\,\rep{2}'')}(T) ~\xrightarrow{\hat{\Sigma}}~ 
j^{(1)}(\hat{\Sigma},T) ~ \rho_{(\rep{1},\,\rep{2}'')}(\hat{\Sigma}) ~ Y^{(1)}_{(\rep{1},\,\rep{2}'')}(T)\;,
\end{equation}
with explicit representation matrices under the $\SG$ and $\TG$ generators $\hat{K}_{\mathrm{S}}$ and $\hat{K}_{\mathrm{T}}$ of $\Gamma'_6$
\begin{equation}
\rho_{(\rep{1},\,\rep{2}'')}(\hat{K}_{\mathrm{S}}) ~=~ \dfrac{-\I}{\sqrt{3}}\begin{pmatrix} 1 & \sqrt{2}\\\sqrt{2}& -1\end{pmatrix} 
\qquad\text{and}\qquad
\rho_{(\rep{1},\,\rep{2}'')}(\hat{K}_{\mathrm{T}}) ~=~ \begin{pmatrix}\omega & 0\\0 & 1\end{pmatrix}\;.
\end{equation}
For the generators of the $S_3$ and $T'$ subgroups of $\Gamma'_6$, defined 
in eqs.\ \eqref{eq:defS3} and \eqref{eq:defTprime}, we get
\begin{subequations}
	\begin{align}
	\rho_{(\rep{1},\,\rep{2}'')}(\mathrm{S}_{S_3}) ~&=~ \Id_2\;, & \rho_{(\rep{1},\,\rep{2}'')}(\mathrm{T}_{S_3}) ~&=~ \Id_2\;,
	\\
	\rho_{(\rep{1},\,\rep{2}'')}(\mathrm{S}_{T'}) ~&=~ \dfrac{-\I}{\sqrt{3}}\begin{pmatrix}1&\sqrt{2}\\\sqrt{2}&-1\end{pmatrix}\;,
	&
	\rho_{(\rep{1},\,\rep{2}'')}(\mathrm{T}_{T'}) ~&=~ \begin{pmatrix}\omega&0\\0&1\end{pmatrix}\;,
	\end{align}
\end{subequations}
such that the modular form transforms as a 
$(\rep{1},\,\rep{2}'')$ representation of $\Gamma'_6\cong S_3 \x T'$.
Note that $Y^{(1)}_{(\rep{1},\,\rep{2}'')}(T)$ is equivalent to the 
weight $1$ modular form of $\Gamma(3)$ and $\mathcal{M}_1(\Gamma(3))\subset\mathcal{M}_1(\Gamma(6))$.
On the other hand, $Y^{(1)}_{(\rep{2},\,\rep{2}')}(T)$ transforms as a 
$(\rep{2},\,\rep{2}')$ representation of $\Gamma'_6$, i.e.\
\begin{equation}
Y^{(1)}_{(\rep{2},\,\rep{2}')}(T) ~\xrightarrow{\hat{\Sigma}}~ 
j^{(1)}(\hat{\Sigma},T) ~ \rho_{(\rep{2},\,\rep{2}')}(\hat{\Sigma}) ~ Y^{(1)}_{(\rep{2},\,\rep{2}')}(T)\;,
\end{equation}
with
\begin{equation} \label{eq:SandTofY1}
\rho_{(\rep{2},\,\rep{2}')}(\hat{K}_\mathrm{S}) = 
\dfrac{-\I}{2\sqrt{3}}
\begin{pmatrix}
-1 & -\sqrt{2} & \phantom{-}\sqrt{3} & \phantom{-}\sqrt{6}\\
-\sqrt{2} & \phantom{-}1 & \phantom{-}\sqrt{6} & -\sqrt{3}\\
\phantom{-}\sqrt{3} & \phantom{-}\sqrt{6} & \phantom{-}1 & \phantom{-}\sqrt{2}\\
\phantom{-}\sqrt{6} & -\sqrt{3} & \phantom{-}\sqrt{2} & -1
\end{pmatrix} 
\quad\text{and}\quad
\rho_{(\rep{2},\,\rep{2}')}(\hat{K}_\mathrm{T}) = 
\begin{pmatrix}
1 & 0 & 0 & 0\\
0 & \omega^2 & 0 & 0\\
0 & 0 & \!\!-1 & 0\\
0 & 0 & 0 & \!\!\!-\omega^2
\end{pmatrix},
\end{equation}
and for the generators of the $S_3$ and $T'$ subgroups we find
\begin{subequations}
	\begin{align}
	\rho_{(\rep{2},\,\rep{2}')}(\mathrm{S}_{S_3}) ~&=~ \dfrac{-1}{2}
	\begin{pmatrix}
	1 & 0 & \sqrt{3} & 0 \\
	0 & 1 & 0 & \sqrt{3} \\
	\sqrt{3} & 0 & -1 & 0 \\
	0 & \sqrt{3} & 0 & -1
	\end{pmatrix}\;, 
	& \rho_{(\rep{2},\,\rep{2}')}(\mathrm{T}_{S_3}) ~&=~ 
	\begin{pmatrix}
	1 & 0 & 0 & 0\\
	0 & 1 & 0 & 0\\
	0 & 0 & \!\!-1 & 0\\
	0 & 0 & 0 & \!\!-1
	\end{pmatrix}\;,
	\\
	\rho_{(\rep{2},\,\rep{2}')}(\mathrm{S}_{T'}) ~&=~ \dfrac{-\I}{\sqrt{3}}
	\begin{pmatrix}
	1 & \sqrt{2} & 0 & 0\\
	\sqrt{2} & -1 & 0 & 0\\
	0 & 0 & 1 & \sqrt{2}\\
	0 & 0 & \sqrt{2} & -1
	\end{pmatrix}\;,
	&
	\rho_{(\rep{2},\,\rep{2}')}(\mathrm{T}_{T'}) ~&=~ 
	\begin{pmatrix}
	1 & 0 & 0 & 0\\
	0 & \omega^2 & 0 & 0\\
	0 & 0 & 1 & 0\\
	0 & 0 & 0 & \omega^2
	\end{pmatrix}\;.
	\end{align}
\end{subequations}
In the decompactification limit, meaning $T\rightarrow\I\infty$, the modular forms go to
\begin{equation}\label{eq:Tiinftylimit_modularforms}
Y^{(1)}_{(\rep{1},\,\rep{2}'')}(T) ~\xrightarrow{T\rightarrow\I\infty}~ (0,1)^\mathrm{T}
\qquad\text{and}\qquad
Y^{(1)}_{(\rep{2},\,\rep{2}')}(T) ~\xrightarrow{T\rightarrow\I\infty}~ (-\sqrt{37},0,0,0)^\mathrm{T}\;,
\end{equation}
as can be seen by combining eqs.\ \eqref{eq:SAGEmodularforms} and \eqref{eq:modularformsZ6}.
Note that eq.~\eqref{eq:SandTofY1} yields $\rho(\hat{K}_\mathrm{S})^2 = -\Id_6$ such that $Y^{(1)}(T)$ 
is invariant under the modular transformation $\mathrm{S}^2=-\Id_2$ of the K\"ahler modulus when 
taking the automorphy factor $(-1)^1$ of $Y^{(1)}(T)$ with weight $1$ into account.

\section{\boldmath The group \texorpdfstring{$[16,13]$}{[16,13]} \unboldmath}
\label{app:16-13}

\begin{table}[!h!]
	\centering
		\begin{tabular}{c|rrrrrrrrrr}
			\toprule
			$[16,13]$  & [Id]  & $[t_2]$ & $[t_3]$ & $[t_2t_3]$ & $[t_1t_3]$ & $[t_1t_2t_3]$ & $[t_1^2]$ & $[t_1^3]$ & $[t_1t_2]$ & $[t_1]$ \\
			\midrule
			$\rep{1}_0$ &$1$&$1$&$1$&$1$&$1$&$1$&$1$&$1$&$1$&$1$\\
			$\rep{1}_1$ &$1$&$-1$&$-1$&$1$&$1$&$-1$&$1$&$-1$&$1$&$-1$\\
			$\rep{1}_2$ &$1$&$-1$&$1$&$-1$&$-1$&$1$&$1$&$-1$&$1$&$-1$\\
			$\rep{1}_3$ &$1$&$-1$&$-1$&$1$&$-1$&$1$&$1$&$1$&$-1$&$1$\\
			$\rep{1}_4$ &$1$&$-1$&$1$&$-1$&$1$&$-1$&$1$&$1$&$-1$&$1$\\
			$\rep{1}_5$ &$1$&$1$&$-1$&$-1$&$1$&$1$&$1$&$-1$&$-1$&$-1$\\
			$\rep{1}_6$ &$1$&$1$&$1$&$1$&$-1$&$-1$&$1$&$-1$&$-1$&$-1$\\
			$\rep{1}_7$ &$1$&$1$&$-1$&$-1$&$-1$&$-1$&$1$&$1$&$1$&$1$\\[4pt]
			$\crep{2}$  &$2$&$0$&$0$&$0$&$0$&$0$&$-2$&$2\I$&$0$&$-2\I$\\
			$\rep{2}$   &$2$&$0$&$0$&$0$&$0$&$0$&$-2$&$-2\I$&$0$&$2\I$\\
			\bottomrule
\end{tabular}
\caption{Character table of the group $[16,13]$.\label{tab:1613char}}
\end{table}

The group $[16,13]\cong(\Z4^{\mathrm{(PG)}}\x\Z2)\rtimes\Z2$ appears as traditional
flavor group of the $\mathbbm{T}^2/\Z4$ orbifold. It can be generated by three generators $t_1,t_2,t_3$,
associated respectively with $\Z4^{\mathrm{(PG)}}$, and the first and second \Z2 factors. They satisfy the presentation
\begin{equation}
   [16,13] ~=~ \left\langle t_1,t_2,t_3 ~|~ t_1^4=t_2^2=t_3^2=t_1^2(t_3t_2)^2=\mathrm{Id},~t_1t_2=t_2t_1,~t_1t_3=t_3t_1\right\rangle\,.
\end{equation}
The character table of this group is given in table~\ref{tab:1613char}.

The tensor products among [16,13] irreducible representations are given by
\begin{subequations}
\begin{align}
\rep1_i\otimes\rep1_i&~=~\rep1_0,\qquad \rep1_0\otimes\rep1_i~=~\rep1_i,\qquad i=0,\ldots,7\\
\rep1_2\otimes\rep1_7&~=~\rep1_3\otimes\rep1_6~=~\rep1_4\otimes\rep1_5~=~\rep1_1,\quad
\rep1_1\otimes\rep1_7 ~=~\rep1_3\otimes\rep1_5~=~\rep1_4\otimes\rep1_6~=~\rep1_2,\\
\rep1_1\otimes\rep1_6&~=~\rep1_2\otimes\rep1_5~=~\rep1_4\otimes\rep1_7~=~\rep1_3,\quad
\rep1_1\otimes\rep1_5~=~\rep1_2\otimes\rep1_6~=~\rep1_3\otimes\rep1_7~=~\rep1_4,\\
\rep1_1\otimes\rep1_4&~=~\rep1_2\otimes\rep1_3~=~\rep1_6\otimes\rep1_7~=~\rep1_5,\quad
\rep1_1\otimes\rep1_3~=~\rep1_2\otimes\rep1_4~=~\rep1_5\otimes\rep1_7~=~\rep1_6,\\
\rep1_1\otimes\rep1_2&~=~\rep1_3\otimes\rep1_4~=~\rep1_5\otimes\rep1_6~=~\rep1_7,\\
\rep1_a\otimes\crep2 &~=~\crep2,\qquad \rep1_a\otimes\rep2~=~\rep2,\qquad a=0,3,4,7,\\
\rep1_a\otimes\crep2 &~=~\rep2,\qquad \rep1_a\otimes\rep2~=~\crep2,\qquad a=1,2,5,6,\\
\rep2\otimes\rep2    &~=~\crep2\otimes\crep2~=~\rep1_1\oplus\rep1_2\oplus\rep1_5\oplus\rep1_6,\qquad
\rep2\otimes\crep2 ~=~\rep1_0\oplus\rep1_3\oplus\rep1_4\oplus\rep1_7\,.
\end{align}
\end{subequations}

\section{\boldmath The modular group \texorpdfstring{$\Gamma_4'$}{Gamma4'} \unboldmath}
\label{app:Gamma4}
The finite modular group $\Gamma_4' \cong S_4' \cong [48,30]$ is presented by
\begin{equation}\label{eq:PresentationGamma4}
\Gamma'_4 ~=~
\left\langle \SG,\,\TG ~\Big|~ \SG^4 ~=~ \TG^4 ~=~ (\SG\,\TG)^3 ~=~ \mathrm{Id} ~,~ \SG^2 \, \TG = \TG \, \SG^2 \right\rangle\;.
\end{equation}
Its tensor products as well as the associated modular forms of $\Gamma(4)$ have been discussed in detail in refs.~\cite{Liu:2020akv,Novichkov:2020eep}.

\subsection{\boldmath Modular binary dihedral group \texorpdfstring{$2D_3$}{2D3}\unboldmath}
\label{sec:2D3}
Of particular interest for the $\T2/\Z4$ orbifold is the binary dihedral
group $2D_3\cong\Z3\rtimes\Z4\cong[12,1]$ and it corresponds to the quotient group
$\Gamma'_4/(\Z2\x\Z2)$ with $\Gamma'_4\cong S_4'$. $2D_3$ is generated by the unfaithful
two-dimensional representation of $\Gamma_4'$ that is denoted as $\rep{\hat{2}}$ in
refs.~\cite{Liu:2020akv,Novichkov:2020eep}. Coincidentally, $2D_3$ is also a subgroup
of $\Gamma'_4$. A presentation for $2D_3$ can be obtained from the presentation of
$\Gamma'_4$ given in eq.~\eqref{eq:PresentationGamma4} by additionally identifying
$\SG^2\TG^2$ with the neutral element, such that
\begin{equation}\label{eq:Presentation2D3}
2D_3 ~=~
\left\langle \SG,\,\TG ~\Big|~ \SG^4 ~=~ \SG^2\TG^2 ~=~ (\SG\,\TG)^3 ~=~ \mathrm{Id},~ \SG^2\TG~=~\TG\SG^2 \right\rangle\,.
\end{equation}
Note that $(\SG^2\TG^2)^2=\TG^4$. The character table of $2D_3$ is given in table \ref{tab:2D3char}.
The tensor products as well as the modular forms of the associated modular subgroup have been given
in ref.~\cite{Arriaga-Osante:2023wnu}. We remark that further modding out the \Z2 generated by $\SG^2$
yields the quotient group $S_3 \cong \Gamma'_4/(\Z2\x\Z2\x\Z2) \cong 2D_3/\Z2 \cong \Gamma_2\cong S_3$,
which is generated by the unfaithful representation $\rep{2}$ of $2D_3$.

\begin{table}[!h!]
	\centering
	\begin{tabular}{l|rrrrrr}
		\toprule
		$2D_3$ &  $[\mathrm{Id}]$ & $[\SG^2]$ & $[\TG]$ & $[\TG\SG]$ & $[\SG]$ & $[\TG\SG^3]$ \\
		\midrule
		$\rep{1}$     & $1$ & $1$ & $1$ & $1$ & $1$ & $1$ \\
		$\rep{1}'$     & $1$ & $1$ & $-1$ & $1$ & $-1$ & $1$\\
		$\rep{\hat{1}}$	& $1$ & $-1$ & $-\I$ & $1$ & $\I$ & $-1$\\
		$\rep{\hat{1}}'$& $1$ & $-1$ & $\I$ & $1$ & $-\I$ & $-1$\\
		$\rep{2}$       & $2$ & $2$ & $0$ & $-1$ & $0$ & $-1$  \\
		$\rep{\hat{2}}$ & $2$ & $-2$ & $0$ & $-1$ & $0$ & $1$  \\
		\bottomrule
	\end{tabular}
	\caption{Character table of $2D_3\cong[12,1]$.}
	\label{tab:2D3char}
\end{table}

\section{\boldmath \texorpdfstring{$\T2/\Z6$}{T2/Z6} 1st twisted sector \unboldmath}
\label{app:Z61st}

In this appendix, we identify the $\Gamma_6'\cong S_3 \x T'$ representation $(\rep{x},\rep{y})$ of
the state $\Phi_{1\mathrm{st}}$ in the 1st twisted sector of the $\T2/\Z6$ orbifold by studying the superpotential
of the low-energy effective theory. Considering the allowed couplings in this orbifold, see e.g.~\cite{Kobayashi:2004ya,Parameswaran:2010ec},
let us take the term
\begin{equation}
\label{eq:WtermPhi1Phi2Phi3}
   \mathcal{W} ~\supset~ Y(T)~\Phi_{1\mathrm{st}} ~ \Phi_{2\mathrm{nd}} ~ \Phi_{3\mathrm{rd}}\;,
\end{equation}
and demand that i) it contains a symmetry-invariant part,
and that ii) there are no unphysical couplings in the decompactification limit,
i.e.\ fields that are infinitely far apart as $T\rightarrow\I\infty$ cannot 
interact with each other. To find the representations $(\rep{x},\rep{y})$ that fulfill the first
criterion, we rewrite eq.~\eqref{eq:WtermPhi1Phi2Phi3} with the explicit components
and representations under the modular flavor symmetry $\Gamma_6'\cong S_3\x T'$ 
of each of the twisted fields and the modular form of weight $1$, i.e.
\begin{equation}
\label{eq:WtermPhi1Phi2Phi3_detail}
  \mathcal{W}_{(\rep{1},\,\rep{1})} \supset
\begin{pmatrix}
Y^{(1)}_{(\rep{2},\,\rep{2}'),1}(T) & Y^{(1)}_{(\rep{2},\,\rep{2}'),2}(T) \\ Y^{(1)}_{(\rep{2},\,\rep{2}'),3}(T) & Y^{(1)}_{(\rep{2},\,\rep{2}'),4}(T)
\end{pmatrix}_{\!\!\!(\rep{2},\,\rep{2}')}
\!\!\!\otimes
\left(\Phi_{1\mathrm{st}}\right)_{(\rep{x},\,\rep{y})}
\otimes
\left(\Phi_{2\mathrm{nd},1},~\Phi_{2\mathrm{nd},2}\right)_{(\rep{1},\,\rep{2}')}
\otimes
\begin{pmatrix}\Phi_{3\mathrm{rd},1}\\\Phi_{3\mathrm{rd},2}\end{pmatrix}_{\!\!\!(\rep{2},\,\rep{1})}
\!\!\!,
\end{equation}
where $S_3$ multiplets are denoted as vertical vectors while components
of $T'$ are arranged horizontally. Here, we have considered the modular weights
of twisted fields given in table~\ref{tab:Z6summary}.
Although eq.~\eqref{eq:modularformsZ6} offers two different vector-valued modular forms
$Y^{(1)}$, $Y^{(1)}_{(\rep{1},\,\rep{2}'')}$ can be disregarded by considering
invariance under $S_3$:
the $S_3$ representation
$\rep{2}$ of $\Phi_{3\mathrm{rd}}$ needs another $\rep{2}$ for the full 
term to contain a trivial singlet representation $\rep{1}$,
cf.~eq.~\eqref{eq:S3ClebschGordans}. Since $\Phi_{1\mathrm{st}}$ and 
hence $\rep{x}$ is a singlet under the modular flavor symmetry, only 
the modular form $Y^{(1)}_{(\rep{2},\,\rep{2}')}(T)$ can provide this 
needed $\rep{2}$ representation of $S_3$.
To determine $\rep{x}$ and $\rep{y}$, we further consider the tensor 
products of $T'$ in eq.~\eqref{eq:WtermPhi1Phi2Phi3_detail}.
For $\mathcal{W}$ to contain a trivial singlet, we must impose
$\rep{1}\supset\rep{y}\otimes(\rep{1}''\oplus\rep{3})$, cf.~eq.~\eqref{eq:TpClebschGordans},
which is only solved by $\rep{y}=\rep{1}'$.
For $\rep{x}$, on the other hand, both $S_3$ representations $\rep{1}$ 
and $\rep{1}'$ would lead to an overall invariant term in the superpotential 
$\mathcal{W}$. Hence, to identify $\rep{x}$, we need to take into account 
the second criterion. In the decompactification limit, where 
$T\rightarrow\I\infty$, all components of the modular form
$Y^{(1)}_{(\rep{2},\,\rep{2}')}(T)$ vanish, except for the first one, 
cf.~eq.~\eqref{eq:Tiinftylimit_modularforms}. Hence, calculating the 
explicit tensor products using eqs.~\eqref{eq:S3ClebschGordans}
and \eqref{eq:TpClebschGordans}, we find that either the interaction
\begin{equation}
Y^{(1)}_{(\rep{2},\,\rep{2}'),1}(T)~\Phi_{1\mathrm{st}}~\Phi_{2\mathrm{nd},2}~\Phi_{3\mathrm{rd},1}
\qquad\text{or}\qquad
Y^{(1)}_{(\rep{2},\,\rep{2}'),1}(T)~\Phi_{1\mathrm{st}}~\Phi_{2\mathrm{nd},2}~\Phi_{3\mathrm{rd},2}
\end{equation}
survives the $T\rightarrow\I\infty$ limit, depending on whether 
$\rep{x}$ is $\rep{1}$ or $\rep{1}'$, respectively. Since 
$\Phi_{1\mathrm{st}}$ and $\Phi_{2\mathrm{nd},2}$ are located 
at the same fixed point as $\Phi_{3\mathrm{rd},2}$, only the second 
interaction is physical such that $\rep{x}=\rep{1}'$.
We conclude that $\Phi_{1\mathrm{st}}$ being a $(\rep{1}',\,\rep{1}')$ 
representation of $\Gamma_6'\cong S_3 \x T'$ is the only admissible representation assignment.

\section{\boldmath Finite modular group of the \texorpdfstring{$\T2/\Z4$}{T2/Z4} 1st twisted sector \unboldmath}
\label{app:Z41stmodular}

In this appendix, we determine the behavior of states in the first twisted
sector of $\T2/\Z4$ under modular transformations,
by using the geometrical automorphisms of the Narain space group and the
Operator Product Expansions (OPEs) for the product of two twisted strings~\cite{Lauer:1990tm}.
Modular transformations arise from the outer automorphisms of the
Narain space group~\cite{Baur:2019iai}, which contain the two generators
\begin{equation}
\hat{K}_\mathrm{S}~=~\begin{pmatrix}0&-\epsilon\\ -\epsilon&0\end{pmatrix} \quad\mathrm{and}\quad \hat{K}_\mathrm{T}~:=~\begin{pmatrix}\Id_2&0\\ -\epsilon&\Id_2 \end{pmatrix}\;,\quad\mathrm{with}\quad\epsilon~=~\begin{pmatrix}0& 1\\ -1 & 0\end{pmatrix}\;,
\end{equation}
that act on the Narain lattice, i.e.\ on the vector 
$(n_1,n_2,m_1,m_2)^\mathrm{T}$, where $n_i$ and $m_i$ are the
winding and Kaluza-Klein numbers, respectively. The behavior of 
untwisted vertex operators $V(\hat{N})$ under such transformations 
has been studied in detail in appendix D.1 of \cite{Baur:2021mtl}. 
There it is noted that they transform as
\begin{equation}
	V(\hat{N}) ~\stackrel{\hat{\Sigma}}{\longrightarrow}~ \varphi_{\hat{\Sigma}}(\hat{N})\, V(\hat{\Sigma}^{-1}\hat{N})\;,
\end{equation}
under $\hat{\Sigma}\in\{\hat{K}_\mathrm{S},\hat{K}_\mathrm{T}\}$, 
where $\varphi_{\hat{\Sigma}}(\hat{N})$ is a phase that originates
from non-commutative effects of closed string operators leading 
to so-called co-cycles. In the $\T2/\Z{4}$ orbifold, 
these states can be gathered into four classes 
of untwisted strings $V^{(n,m)}$ with $n,m\in\{0,1\}$ 
and $n=\sum n_i\mod2$, $m=\sum m_i \mod2$, which 
transform under the modular symmetry as
\begin{subequations}
	\label{eq:UntwistedUnderModular}
	\begin{align}
		V^{(0,0)} ~\stackrel{\hat{K}_\mathrm{S}}{\longrightarrow}~ V^{(0,0)}\;, && V^{(1,1)} ~\stackrel{\hat{K}_\mathrm{S}}{\longrightarrow}~ -V^{(1,1)}\;, && V^{(1,0)} ~\stackrel{\hat{K}_\mathrm{S}}{\longleftrightarrow}~ V^{(0,1)}\;,\\
		V^{(0,0)} ~\stackrel{\hat{K}_\mathrm{T}}{\longrightarrow}~ V^{(0,0)}\;, && V^{(0,1)} ~\stackrel{\hat{K}_\mathrm{T}}{\longrightarrow}~ -V^{(0,1)}\;, && V^{(1,0)} ~\stackrel{\hat{K}_\mathrm{T}}{\longleftrightarrow}~ V^{(1,1)}\;.
	\end{align}
\end{subequations}
Note that, despite the additional phases, the untwisted strings $V^{(n,m)}$ 
transform trivially under $\hat{K}_\mathrm{S}^2$, $\hat{K}_\mathrm{T}^2$, 
and $(\hat{K}_\mathrm{T}\hat{K}_\mathrm{S})^3$.
The OPEs for two twisted strings of the first and third twisted sector
of the $\T2/\Z4$ orbifold have been determined in ref.~\cite{Lauer:1990tm}.
By inverting them, one obtains
\begin{subequations}
	\begin{align}
		V^{(0,0)} ~&=~ \bar{\phi}_1\phi_1 + \bar{\phi}_2 \phi_2\;,\\
		V^{(1,0)} ~&=~ \bar{\phi}_2\phi_1 + \bar{\phi}_1 \phi_2\;,\\
		V^{(0,1)} ~&=~ \bar{\phi}_1\phi_1 - \bar{\phi}_2 \phi_2\;,\\
		V^{(1,1)} ~&=~ -\I\, \bar{\phi}_2\phi_1 +\I\, \bar{\phi}_1 \phi_2\;,
	\end{align}
\end{subequations}
where $\Phi_{1\mathrm{st}}=(\phi_1,\phi_2)^\mathrm{T}$ 
and $\Phi_{3\mathrm{rd}}=(\bar{\phi}_1,\bar{\phi}_2)^\mathrm{T}$.
From these OPEs together with the behavior of the untwisted strings under the
modular symmetry, given in eq.~\eqref{eq:UntwistedUnderModular}, we obtain
that the twisted strings transform under the modular symmetry as
\begin{equation}
	\Phi_{1\mathrm{st}} ~\stackrel{\hat{K}_\mathrm{S}}{\longrightarrow}~ 
	\dfrac{\I^a}{\sqrt{2}}\begin{pmatrix}1&1\\1&-1\end{pmatrix}\Phi_{1\mathrm{st}}\;,
	\qquad\text{and}\qquad
	\Phi_{1\mathrm{st}} ~\stackrel{\hat{K}_\mathrm{T}}{\longrightarrow}~ 
	\begin{pmatrix}0&\e^{2\pi\I\,b}\\\e^{2\pi\I\,(b+\nicefrac{3}{4})}&0\end{pmatrix}\Phi_{1\mathrm{st}}\;,
\end{equation}
with $a\in\Z{}$ and $b\in\mathbbm{R}$. 
By demanding that the trivial modular transformation
$(\hat{K}_\mathrm{T}\hat{K}_\mathrm{S})^3$ acts trivially on the 
twisted strings, we obtain $a=0$, $b=\nicefrac{5}{8}$.
We remark that the additional solution $a=2$, $b=\nicefrac{1}{8}$ only
differs by a multiplication with the negative identity matrix, which 
becomes irrelevant in the effective field theory, since its action only 
contains terms that have an even amount of fields from the first or third 
twisted sectors. Thus, we arrive at
\begin{equation}\label{eq:TrafoPhi1st_OPE_app}
	\rho_\mathrm{1st}(\hat K_\mathrm{S}) = \frac{1}{\sqrt2}
	\begin{pmatrix}
	1 & 1\\
	1 & -1
	\end{pmatrix},
	\qquad
	\rho_\mathrm{1st}(\hat K_\mathrm{T}) =
	\begin{pmatrix}
	0 & \e^{\nicefrac{2\pi\I5}{8}}\\
	\e^{\nicefrac{2\pi\I3}{8}} &  0 
	\end{pmatrix},
\end{equation}
which builds the unfaithful representation $\rep{2}$ of
$2D_3\cong[12,1]$ that only spans an $S_3$.

{\small
\providecommand{\bysame}{\leavevmode\hbox to3em{\hrulefill}\thinspace}

}
\end{document}